\documentclass[amsmath,superscriptaddress,aps,prb,twocolumn]{revtex4-1}
\usepackage[colorlinks,linkcolor=blue,anchorcolor=blue,citecolor=blue,urlcolor=blue]{hyperref}
\usepackage{amsmath}
\usepackage{amssymb}
\usepackage{graphicx}
\usepackage{color}
\usepackage{bm}

\begin{document}
\title{Floquet topological properties in the Non-Hermitian long-range system with complex hopping amplitudes }

\author{Gang-Feng Guo}
\thanks{These authors contributed equally to this work.}
\affiliation{Lanzhou Center for Theoretical Physics, Key Laboratory of Theoretical Physics of Gansu Province, Lanzhou University, Lanzhou $730000$, China}

\author{Yan Wang}
\thanks{These authors contributed equally to this work.}
\affiliation{Cuiying Honors College, Lanzhou University, Lanzhou $730000$, China}

\author{Xi-Xi Bao}
\affiliation{Lanzhou Center for Theoretical Physics, Key Laboratory of Theoretical Physics of Gansu Province, Lanzhou University, Lanzhou $730000$, China}

\author{Lei Tan}
\email{tanlei@lzu.edu.cn}
\affiliation{Lanzhou Center for Theoretical Physics, Key Laboratory of Theoretical Physics of Gansu Province, Lanzhou University, Lanzhou $730000$, China}
\affiliation{Key Laboratory for Magnetism and Magnetic Materials of the Ministry of Education, Lanzhou University, Lanzhou $730000$, China}

\begin{abstract}

Non-equilibrium phases of matter have attracted much attention in recent years, among which the Floquet phase is a hot point. In this work, based on the Periodic driving Non-Hermitian model, we reveal that the winding number calculated in the framework of the Bloch band theory has a direct connection with the number of edge states even the Non-Hermiticity is present. Further, we find that the change of the phase of the hopping amplitude can induce the topological phase transitions. Precisely speaking, the increase of the value of the phase can bring the system into the larger topological phase. Moreover, it can be unveiled that the introduction of the purely imaginary hopping term brings an extremely rich phase diagram. In addition, we can select the even topological invariant exactly from the unlimited winding numbers if we only consider the next-nearest neighbor hopping term. Here, the results obtained may be useful for understanding the Periodic driving Non-Hermitian theory.

\end{abstract}

\maketitle
\section{INTRODUCTION}\label{I}

In quantum mechanics, the physical observable is represented by the Hermitian operator which has the real eigenvalues [\onlinecite{1}]. However, in the real world, the system will inevitably be coupled with the environment, and the Non-Hermitian Hamiltonian is necessary to describe the systems naturally, such as the wave propagations with gain and loss [\onlinecite{11,12,13,14,15,16,17,18,19,20,21+1,21}] and various open systems [\onlinecite{3,4,5,6,7,8,9,10}]. For simplicity, we usually take the hopping amplitudes to be real [\onlinecite{21,3,4,5,6,7,8,9,10,23,24,25,26,27,28,29,30,31,32,33}]. However, sometimes, we need to take the phase of hopping terms into account since the phase itself is no longer trivial [\onlinecite{34,35,36,37,38,39,40,41,43,44,46}]. The introduction of the phase of hopping terms, or equivalently, the complex hopping terms can not only cause the Non-Hermiticity of the system, but also exhibit rich topological phases [\onlinecite{34,35,36,37,38,39,40,41,43,44,46}]. For example, Ref. [\onlinecite{41}] has introduced complex next-nearest-neighbour tunneling terms to break time-reversal symmetry and finally realized the quantum Hall effect on a honeycomb lattice. Ref. [\onlinecite{44}] introduced a purely imaginary next-nearest-neighbor hopping in the Kane-Mele-Hubbard model which holds a particle-hole symmetry, resulting in the absence of the charge and spin currents and the absence of the quantum Monte Carlo sign problem. Refs. [\onlinecite{38}, \onlinecite{46}] found that vortex-solid states exist at $f = p/q$ where $f$ is the magnetic flux quanta per the fundamental plaquette and $p$ and $q$ are co-prime natural numbers.

On the other hand, the long-range hopping has been noticed in many systems [\onlinecite{28,64,65,66,67,68,69,70,70-1}]. Related researches showed that long-range hopping can change the symmetry classification and topological invariants [\onlinecite{69}]. The fractional topological number can also appear in the one-dimensional Non-Hermitian long-range lattice system [\onlinecite{70}]. In addition, the long-range Su-Schrieffer-Heeger (SSH) model can exhibit a topological phase diagram that contains not only the winding numbers $w = 0$, $1$, but $2$ [\onlinecite{70-1}, \onlinecite{701}]. Similar to the effect of the long-range hopping, the topological phases with large topological numbers also can be obtained by the Periodic driving. As we all know, Periodic driving has been used as a powerful control tool to achieve coherent control of quantum states [\onlinecite{47,48,49,50,51}], and artificially create exotic topological phases in systems of ultracold atoms [\onlinecite{53,54,55,atom1,atom2,atom3}], photonics [\onlinecite{56}, \onlinecite{57}], superconductor qubits [\onlinecite{58}, \onlinecite{qubit}] and graphene [\onlinecite{59,gra1,gra2,gra3}]. Periodic driven engineering can not only realize topological phases that are difficult to achieve in static systems, but also provide effective control of topological phase transitions [\onlinecite{an1,an2,zhou1,zhou2}], to realize many topological states that do not exist in static systems [\onlinecite{60,61,62,63}]. To date, we have known that the effect of long-range hopping and the Periodic driving on the system, respectively. However, it is still a secret for us that what will happen if we combine the long-range hopping and the Periodic driving, and further, what new phenomena will arse if the hopping amplitudes become complex especially in the Non-Hermitian cases? As we will clarify in the next, the combination of these ingredients will endow the system with more novel phenomena.

Concretely, we will discuss a Non-Hermitian Floquet long-range model with complex hopping amplitudes. The main results are as follows: (i) In this system, there is a direct connection between the winding numbers determined by the Bloch theory and the number of edge state pairs. (ii) The introduction of the purely imaginary hopping term enriches the phase diagram, making it no longer as monotonous as the phase diagram in the real number field. (iii) The phase of the hopping term increases within a certain range, making the system transform to a topological phase with a large topological number. (iv) The long-range hopping term plays a role in selecting an even topological number from an unlimited winding numbers in complex number induced Non-Hermitian Floquet topological system.

The paper is organized as follows: The theoretical framework is constructed in Sec. \ref{II}. Sec. \ref{III} focuses on the topological properties of edge states and reveals the bulk-boundary correspondence in the model. In Sec. \ref{IV}, we introduce the complex-valued hopping term and the long-range hopping to discuss the influence on topological properties. Finally, the conclusion is given in Sec. \ref{V}.

\section{MODEL AND THEORY}\label{II}

In this paper, we focus on a Non-Hermitian long-range model, which is subjected to piecewise time-periodic quenches and the time-dependent Hamiltonian has the following form:

\begin{eqnarray}
\hat H(t) = \left\{ \begin{array}{ll}
\hat H_{1} &    for (m-1)T<t<(m-\frac{1}{2})T\\
\hat H_{2} &    for (m-\frac{1}{2})T<t<mT\\
\end{array} \right.,\label{1}
\end{eqnarray}

in which
\begin{eqnarray}
\hat H_{1}=\sum_{n} [&[{\hat c^\dag_{n}}\hat \sigma^{+}(t_{1}\hat c_{n+1}+t_{2}\hat c_{n+2})+(t_{1}\hat c^\dag_{n+1}+t_{2}\hat c^\dag_{n+2})\hat \sigma^{+}\hat c_{n}\nonumber\\
&+H.c.]+2\gamma(\hat c^\dag_{n}\hat \sigma^{+}\hat c_{n}-\hat c^\dag_{n}\hat \sigma^{-}\hat c_{n})]\label{2},
\end{eqnarray}
and
\begin{eqnarray}
\hat H_{2}=\sum_{n} [&(\omega_{1}\hat c^\dag_{n+1}+\omega_{2}\hat c^\dag_{n+2})\hat \sigma^{+}\hat c_{n}-\hat c^\dag_{n}\hat \sigma^{+}(\omega_{1}\hat c_{n+1}+\omega_{2}\hat c_{n+2})\nonumber\\
&+2\mu\hat  c^\dag_{n}\hat \sigma^{+}\hat c_{n}+H.c.].\label{3}
\end{eqnarray}
Where $\hat c^\dag_{n}=(\hat c^\dag_{n,A},\hat c^\dag_{n,B})$ and $\hat \sigma^{\pm}=\frac{\hat \sigma_{x}\pm i\hat \sigma_{y}}{2}$. $l\in \mathbb{Z}$,  $T$ is the driving period and $T=1$. $c^\dag_{n,A(B)}(\hat c_{n,A(B)})$ is the creation (annihilation) operator at the A(B) sublattice site on the $nth$ unit cell. In each driving period, we choose a generalized SSH model. In the first half of a driving period, the lattice Hamiltonian contains intercell hoppings $t_{1}$, $t_{2}$ and intracell asymmetric hopping amplitudes $\pm\gamma$ which can introduce Non-Hermitian effects. In the second half of a driving period, the lattice is a long range SSH model, the Hamiltonian contains intercell hoppings $\omega_{1}$, $\omega_{2}$ and intracell hopping amplitude $\mu$. Clearly, the Hamiltonian $\hat H_{2}$ maintains Hermitian. $t_{1}$ and $\omega_{1}$ represent the nearest neighbor hopping, $t_{2}$ and $\omega_{2}$ denote the long-range hopping.

To clarify the topological properties of the system, the topological invariant should be considered. In momentum space, the Floquet operator $\hat U$ describing the evolution of the system can be defined as

\begin{eqnarray}
\hat U(k)=e^{-i \frac{\hat H_{2}(k)}{2}}e^{-i \frac{\hat H_{1}(k)}{2}},\label{4}
\end{eqnarray}
where
\begin{eqnarray}
\hat H_{1}(k)=2i\gamma\hat \sigma_{y}+(2t_{1}\cos(k)+2t_{2}\cos(2k))\hat \sigma_{x},\label{5}
\end{eqnarray}
and
\begin{eqnarray}
\hat H_{2}(k)=2\mu\hat \sigma_{x}+(2\omega_{1}\sin(k)+2\omega_{2}\sin(2k))\hat \sigma_{y}.\label{6}
\end{eqnarray}

By construction, $k\in(-\pi, \pi]$ is the quasimomentum. We note that the Floquet operator $\hat U(k)$ defined above has sublattice symmetry in two symmetric time frames [\onlinecite{71},\onlinecite{72}]. These frames are obtained by shifting the starting time of the evolution forward and backward, respectively [\onlinecite{63}]. The resulting Floquet operators are given by

\begin{eqnarray}
\hat U_{1}(k)=e^{-i \frac{\hat H_{2}(k)}{4}}e^{-i \frac{\hat H_{1}(k)}{2}}e^{-i \frac{\hat H_{2}(k)}{4}},\label{7}
\end{eqnarray}
and
\begin{eqnarray}
\hat U_{2}(k)=e^{-i \frac{\hat H_{1}(k)}{4}}e^{-i \frac{\hat H_{2}(k)}{2}}e^{-i \frac{\hat H_{1}(k)}{4}}.\label{8}
\end{eqnarray}

Obviously, $\hat U_{1}(k)$ and $\hat U_{2}(k)$ come from the similarity transformation of $\hat U(k)$, and therefore have the same complex quasienergy spectrum. It follows that $\hat U_{1}(k)$ and $\hat U_{2}(k)$ possess a sublattice symmetry defined by the operator $\hat \sigma_{z}$, such that $\hat \sigma_{z} \hat U_{1,(2)}(k)\hat \sigma_{z}^{-1}=\hat U_{1,(2)}^{-1}(k)$.

With the sublattice symmetry, the winding number for $\hat U_{1}(k)$ and $\hat U_{2}(k)$ can be defined using the method in Ref. [\onlinecite{63}].
\begin{equation}
W_{s}=\int_{ - \pi }^{ + \pi } \frac{dk}{2\pi} \frac{n_{sx}(k)\partial_{k}n_{sy}(k)-n_{sy}(k)\partial_{k}n_{sx}(k)}{n^{2}_{sx}(k)+n^{2}_{sy}(k)},\label{9}
\end{equation}
for $s=1,2$, in which $n_{sx}(k)$ and $n_{sy}(k)$ are elements of $\hat U_{s}(k)$
\begin{equation}
\hat U_{s}(k)=\cos[E(k)]\hat I-i[n_{sx}(k)\hat \sigma_{x}+n_{sy}(k)\hat \sigma_{y}],\label{10}
\end{equation}
and the components $n_{sx}(k)$ and $n_{sy}(k)$ are given by
\begin{equation}
n_{1x}(k) = \sin[h_{x}(k)]\cos[h_{y}(k)+i\gamma],\label{11}
\end{equation}
\begin{equation}
n_{1y}(k) = \sin[h_{y}(k)+i\gamma],\label{12}
\end{equation}
\begin{equation}
n_{2x}(k) = \sin[h_{x}(k)],\label{13}
\end{equation}
\begin{equation}
n_{2y}(k) = \cos[h_{x}(k)]\sin[h_{y}(k)+i\gamma],\label{14}
\end{equation}
in which
\begin{equation}
h_{x}(k) = \mu+t_{1}\cos(k)+	t_{2}\cos(2k),\label{15}
\end{equation}
\begin{equation}
h_{y}(k) = \omega_{1}\sin(k)+	\omega_{2}\sin(2k).\label{16}
\end{equation}

We note that although $n_{sx}(k)$ and $n_{sy}(k)$ have Non-Hermiticity, the definite integration is still a real integer [\onlinecite{73}]. Combining these winding numbers, the complete topological characterization of the system  can be introduced as
\begin{equation}
W_{0}=\frac{W_{1}+W_{2}}{2},\label{17}
\end{equation}
\begin{equation}
W_{\pi}=\frac{W_{1}-W_{2}}{2}.\label{18}
\end{equation}
These winding numbers will be used to characterize the Non-Hermitian Floquet topological phases in our system.

\section{Edge states properties}\label{III}
\begin{figure}[!htbp]
\includegraphics[width=4.2cm,height=3.8cm]{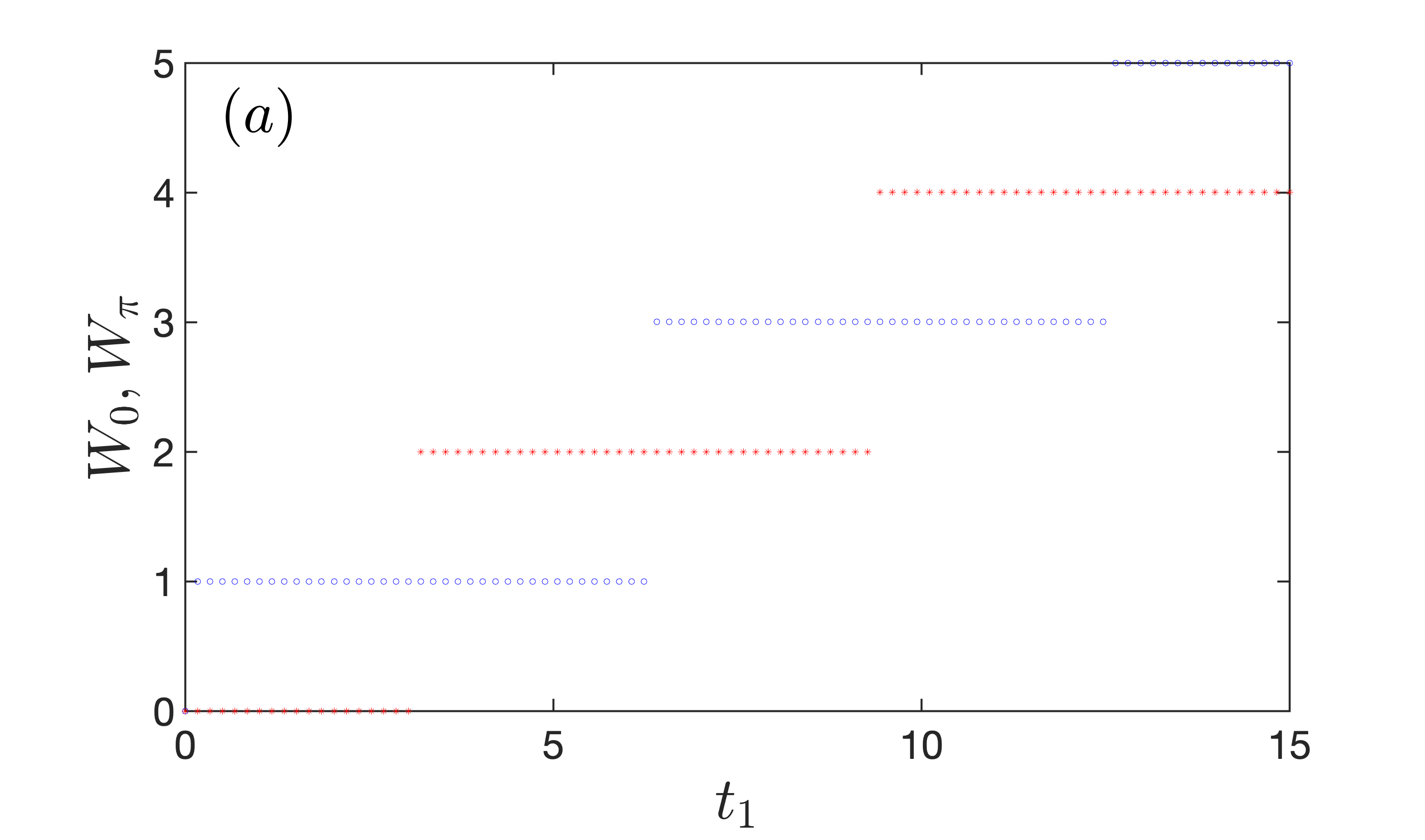}
\includegraphics[width=4.2cm,height=3.8cm]{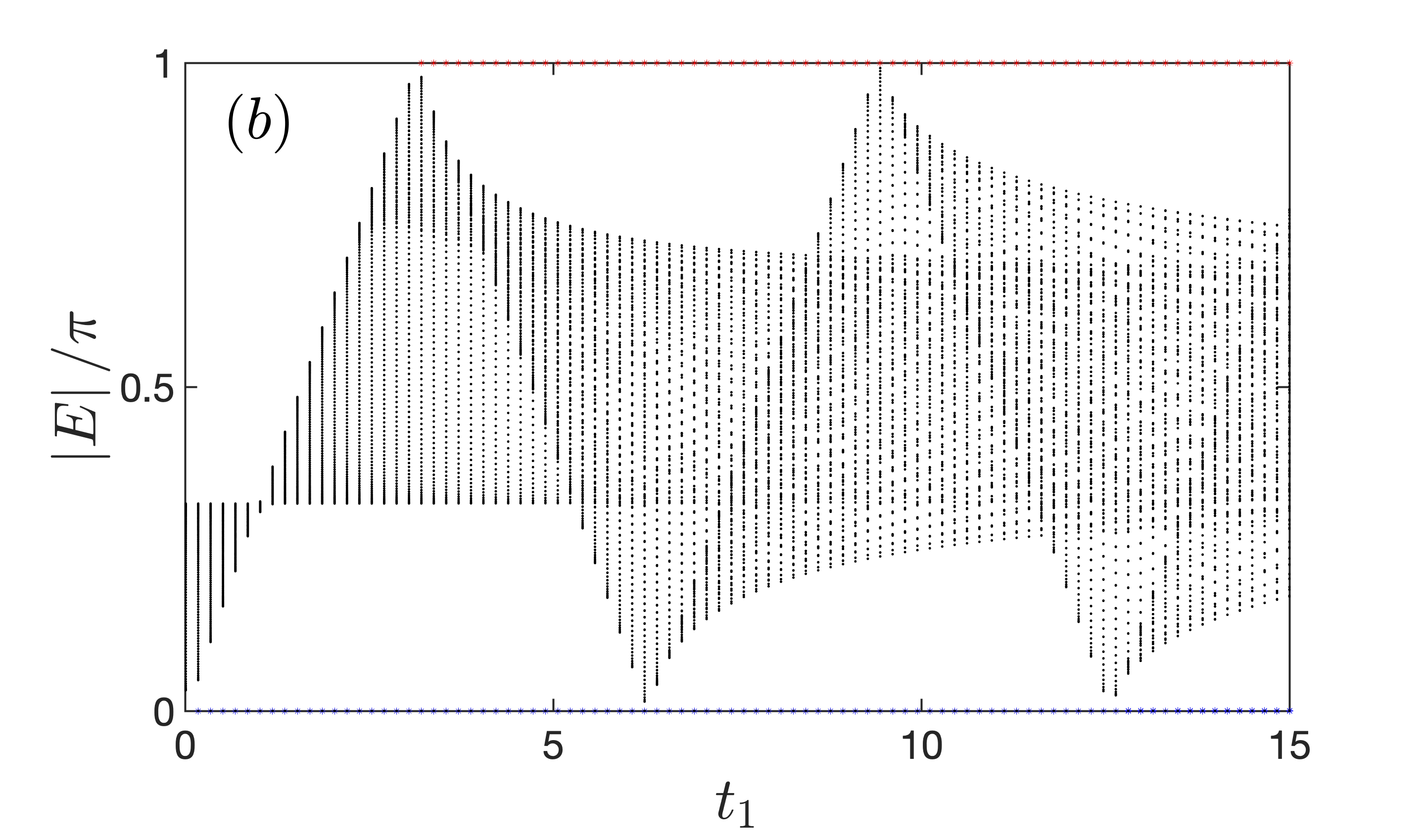}
\includegraphics[width=4.2cm,height=3.8cm]{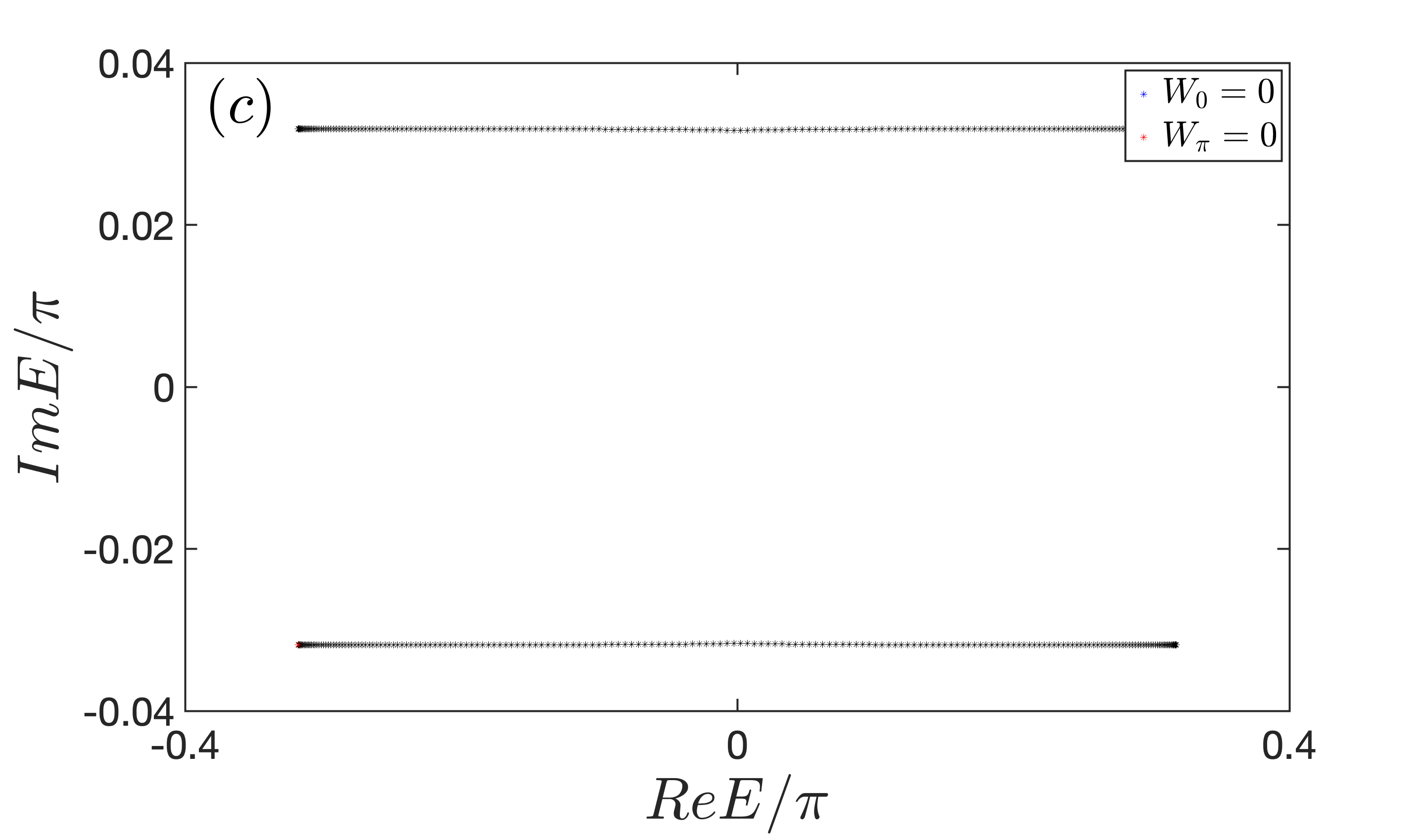}
\includegraphics[width=4.2cm,height=3.8cm]{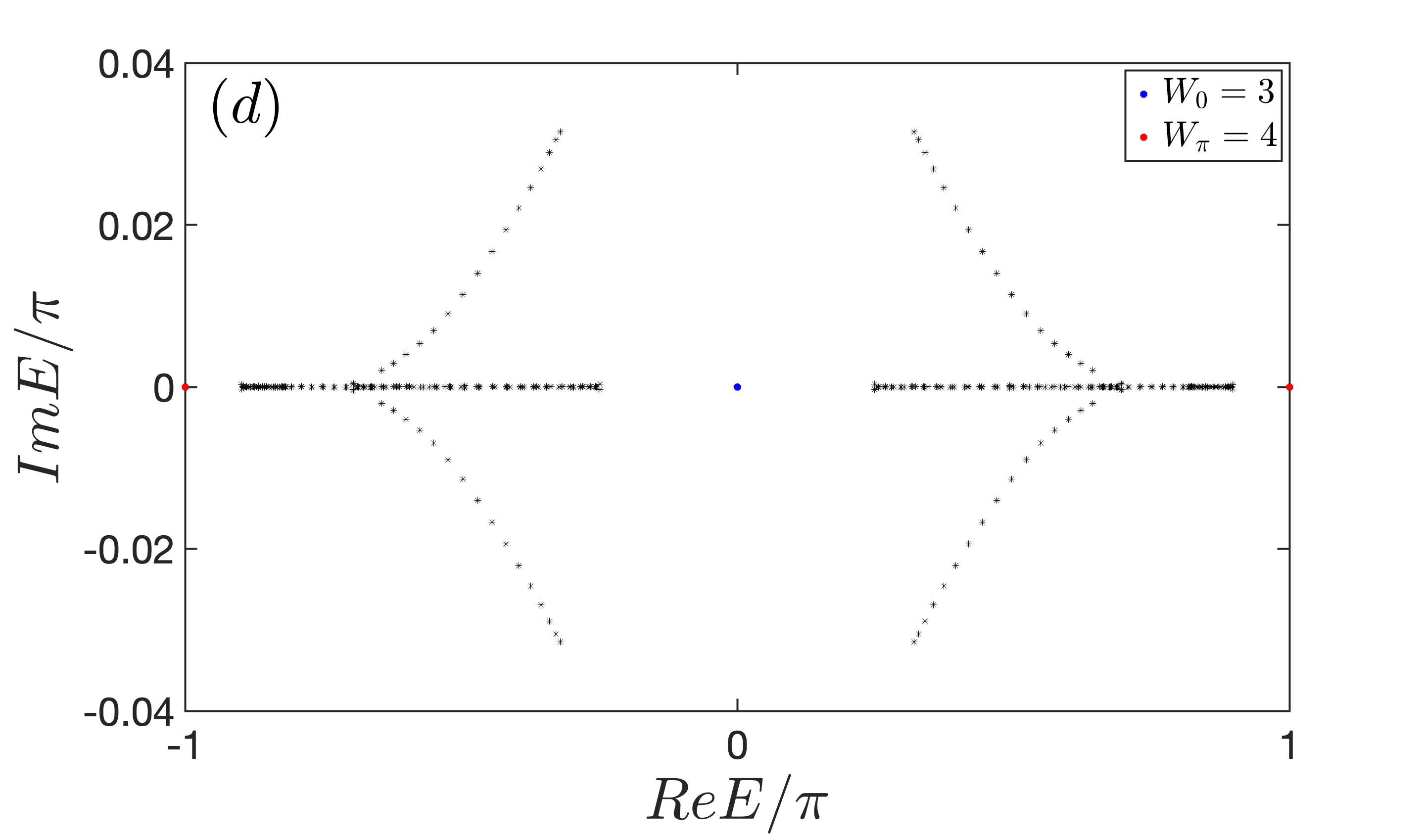}
\caption{(Color online) (a) A pair of winding numbers $W_{0}$ (blue) and $W_{\pi}$(red) versus system parameters $t_{1}$. (b) The quasienergy spectrum of the Floquet operator $\hat U$ in open boundary condition. The blue dots denote the multi-degenerate zero modes of the open chain while the red dots stand for the multi-degenerate $\pi$ modes. (c) $t_{1}=0.01$,  no edge states appears. (d) $t_{1}=10$, which corresponds to three pairs of edge states at quasienergy $E=0$ and four  pairs of edge states at quasienergy $E=\pm\pi$. The common parameters are given by $N=200$, $\gamma=0.1$, $t_{2}=0$, $\mu=0$, $\omega_{1}=1$ and $\omega_{2}=0$. }
\label{fig1}
\end{figure}

To reveal the topological properties of the model, the energy spectrum and topological invariant can be depicted in Fig. \ref{fig1}. All parameters are taken as real numbers for simplicity in this section.

We show in Fig. \ref{fig1}(a) the calculations of a pair of winding numbers $W_{0}$ (blue) and $W_{\pi}$(red) versus the system parameter $t_{1}$. We can find that the winding numbers only take integer values, as suggested by the above theory. When the value of the system parameter $t_{1}$ is not restricted, the $0-$mode and $\pi-$mode states in the quasi-energy spectrum can be generated through Periodic driving engineering, which can artificially synthesize topological phases with large topological numbers [\onlinecite{63}]. We also present the absolute values of quasienergy $E$ versus the amplitude $t_{1}$ under open boundary condition in Fig. \ref{fig1}(b). Clearly, the multi-degenerate $0-$mode and $\pi-$mode states appear.

Then we find that edge states at $E = 0$ and $\pm\pi$ are surrounded by gaps in the complex quasienergy plane $ReE-ImE$ and have real quasienergiess as shown in Figs. \ref{fig1}(c)-(d). There is no edge states at quasienergy $E=0$ and $\pm\pi$ when $t_{1}=0.01$ as shown in Fig. \ref{fig1}(c). While there are three pairs of edge states at quasienergy $E=0$ and four pairs of edge states at quasienergy $E=\pm\pi$ when $t_{1}=10$ as shown in Fig. \ref{fig1}(d). Through the calculation of Eq. \eqref{9}, we find that the absolute value of the winding number $W_{0}$ and $W_{\pi}$ are equal to the number of edge state pairs at quasienergies zero and $\pi$, that is, $n_{0/\pi}=\left| W_{0/\pi} \right|$. In other words, the winding number of the bulk can be used to predict the number of edge states, which is the bulk-boundary correspondence in the Hermitian system [\onlinecite{71},\onlinecite{72}].

\begin{figure}[!htbp]
\includegraphics[width=4.2cm,height=3.8cm]{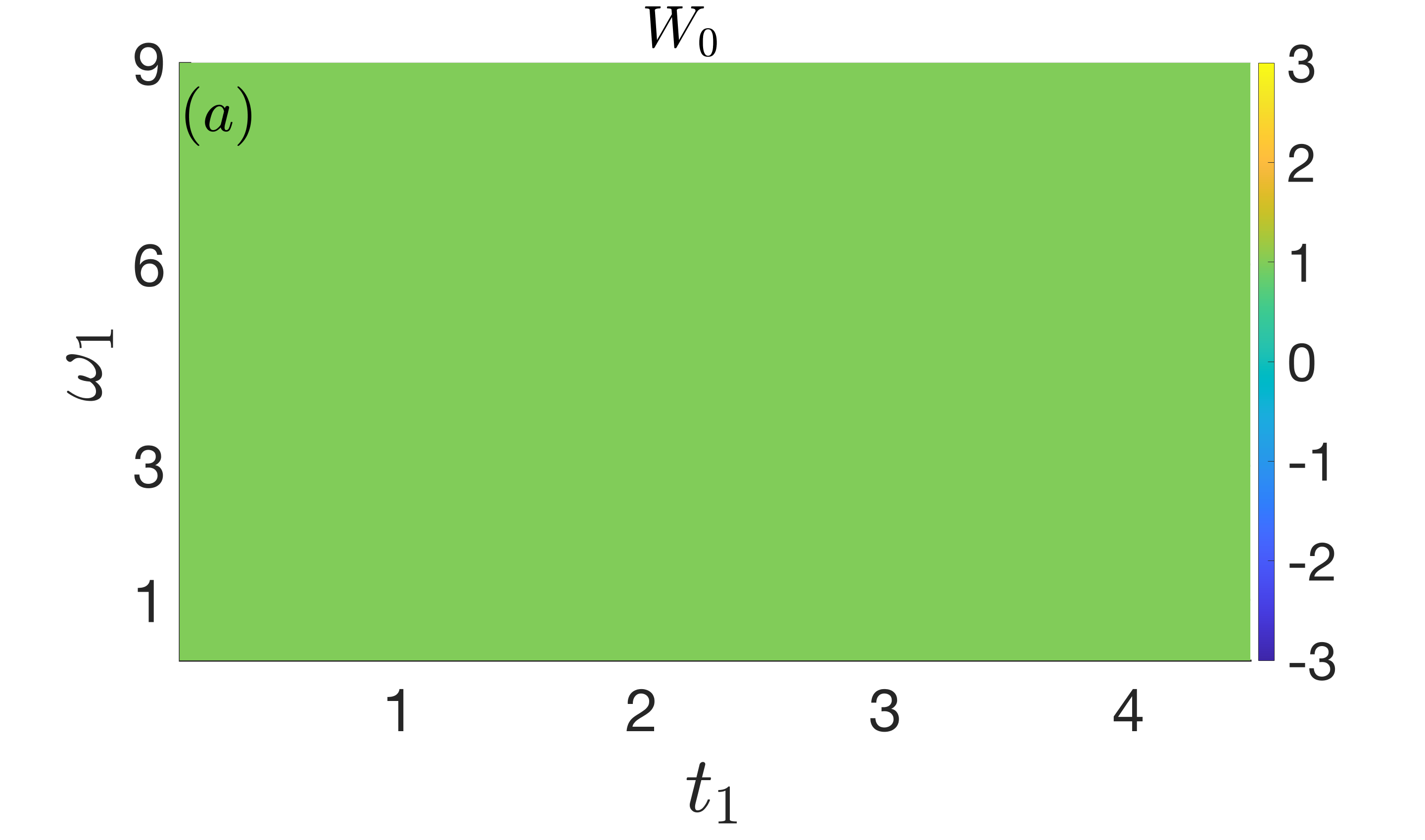}
\includegraphics[width=4.2cm,height=3.8cm]{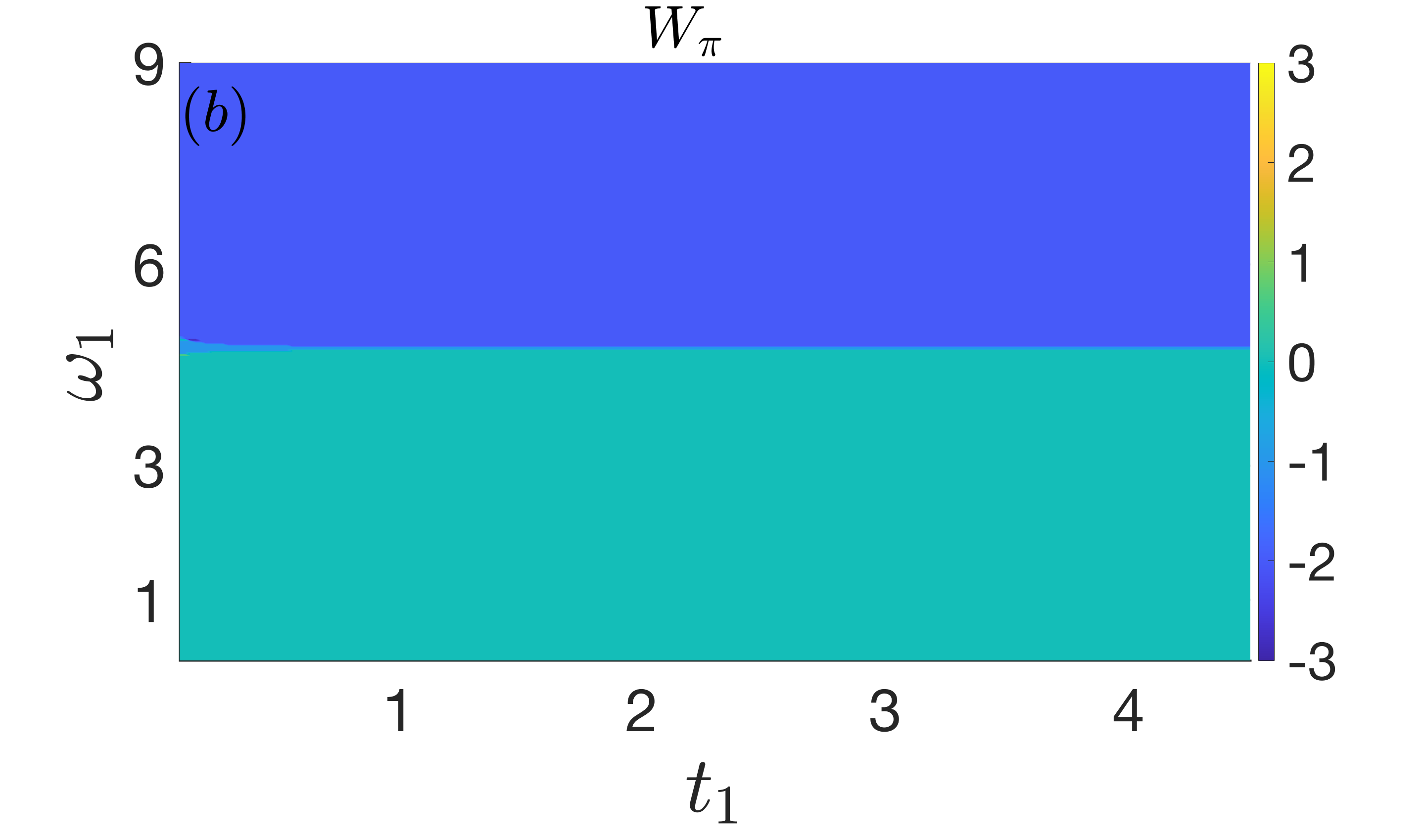}
\includegraphics[width=4.2cm,height=3.8cm]{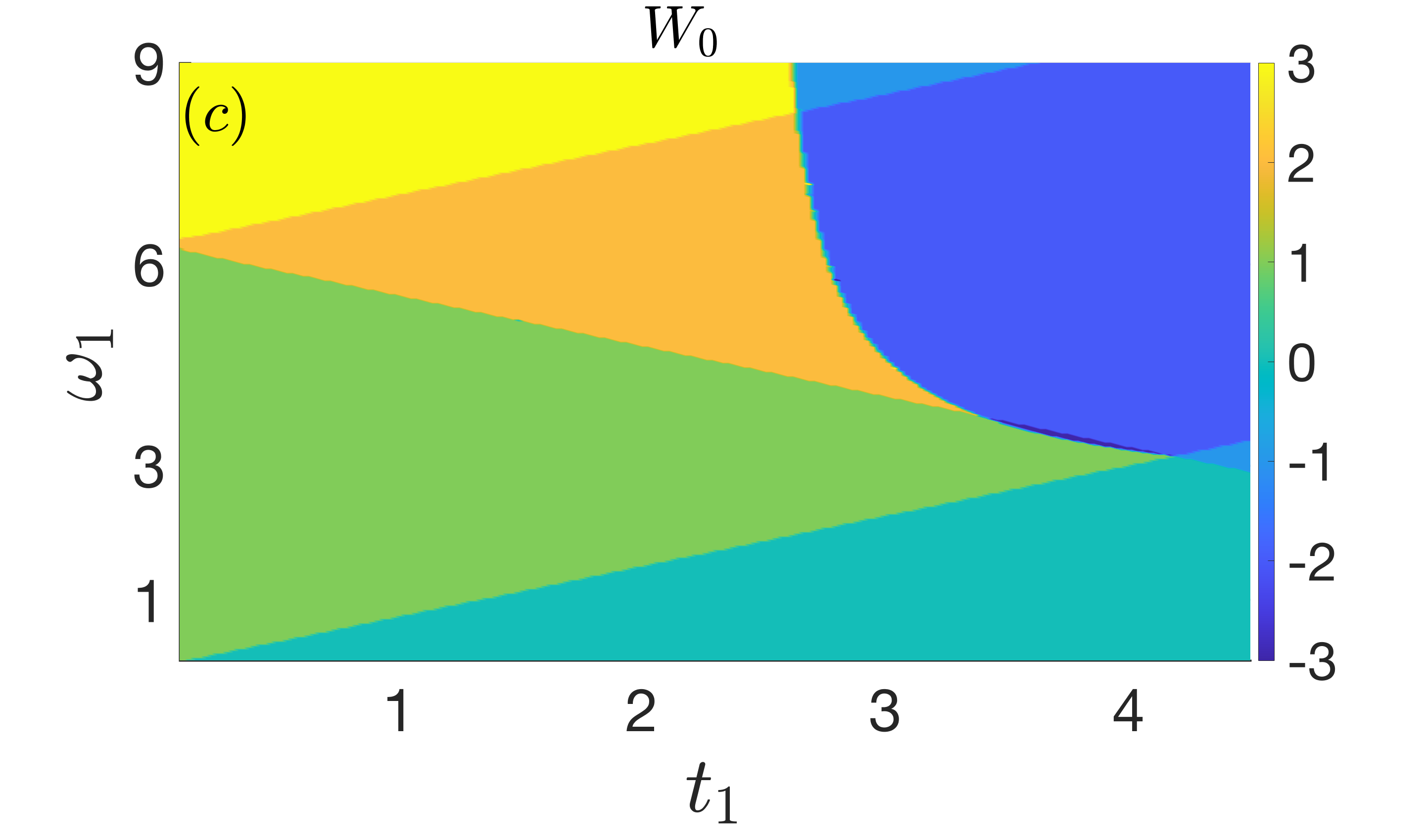}
\includegraphics[width=4.2cm,height=3.8cm]{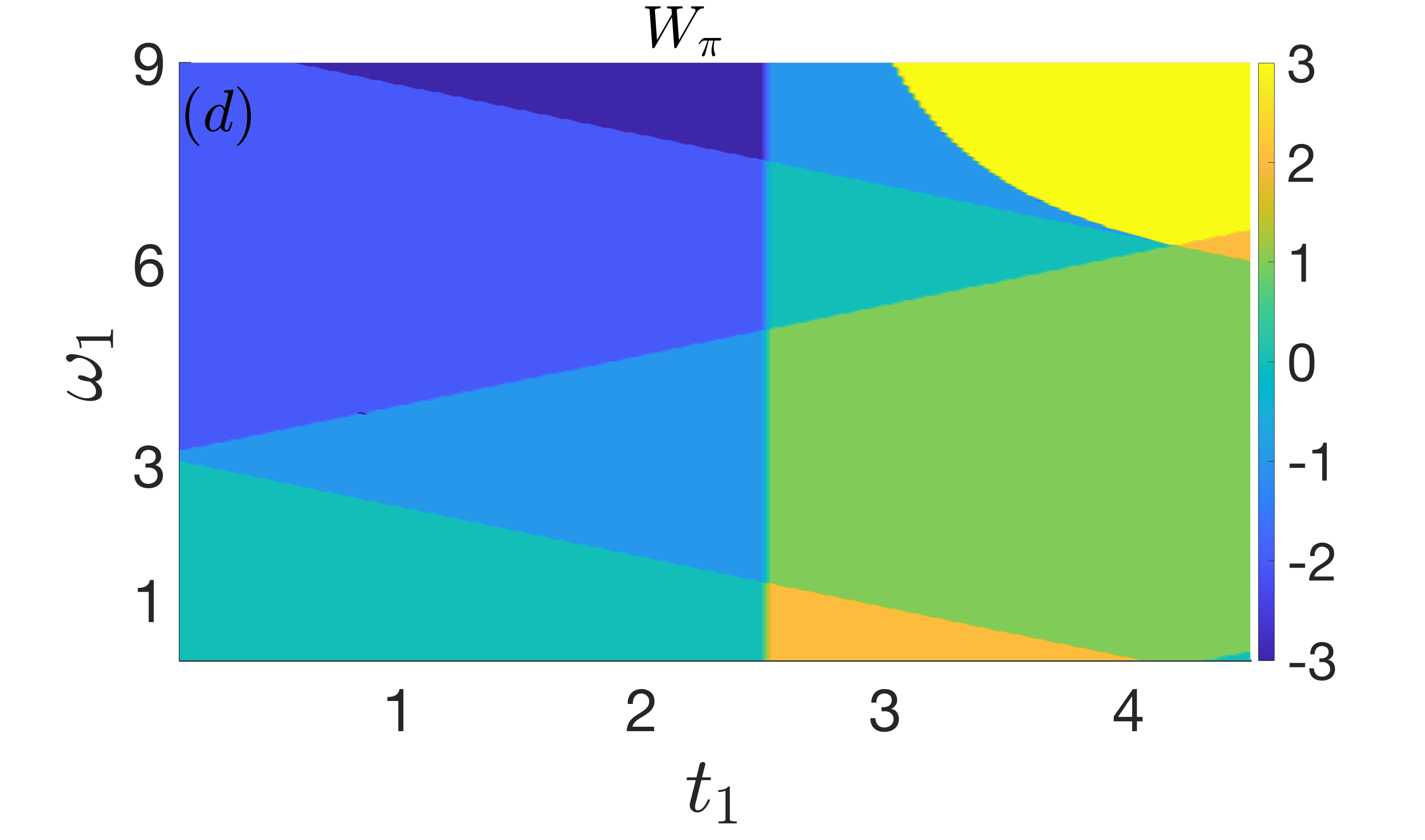}

\caption{(Color online) A pair of winding numbers with system hopping amplitudes $t_{1}$ and $\omega_{1}$, the common parameters are given by $N=200$, $t_{2}=0.01$, $\mu=0$ and $\omega_{2}=0.01$. For (a)-(b) $\gamma = 0.75t_{1}$. For (c)-(d) $\gamma = 0.75it_{1}$.}
\label{fig2}
\end{figure}

\section{non-Hermiticity induced by Complex number}\label{IV}

In this section, we discuss the topological properties of this system when considering the phase of the hopping. We first choose the intracell asymmetric hopping amplitude $\gamma=0.75t_{1}$, all the parameters are real numbers, the topological invariant can be displayed in Figs. \ref{fig2}(a)-(b). Clearly, the value of $W_{0}$ is always 1 if the Non-Hermiticity takes the real number. In contrast, the $\pi$ modes are not all 0, but it can also take -2 and -1. As we can see, these values are relatively simple, and there is no very complicated distribution. However, this situation becomes very different if we take the phases of the hopping amplitudes into account.

When we consider complex hopping amplitudes with phases, intracell hopping amplitude $\gamma=\gamma_{0} e^{i\theta}$, $\theta\in(0, 2\pi]$ where $\theta$ is the relative phase with respect to $t_{1}$ and the phases of $t_{1}$ and other parameters are all set to 0. A special case is $\theta=\frac{\pi}{2}$, $\gamma=0.75i*t_{1}$, which is a pure imaginary number. Then we show the topological phase diagram of the system under this special phase. In Figs. \ref{fig2}(c)-(d), the system parameter $\gamma=0.75i*t_{1}$, which will cause the system to degenerate from Non-Hermitian to Hermitian, since the complex conjugate of $i$ is $-i$. The pure imaginary asymmetric term $\gamma$ on the off-diagonal of the Hamiltonian will become its opposite when performing complex conjugate calculations, and then become the diagonal term of the initial Hamiltonian when the transpose operation is processed, that is, $\hat H_{1}(k)=\begin{pmatrix} 0 & 2\gamma+\triangle \\ -2\gamma+\triangle & 0 \end{pmatrix}\\$ $\Rightarrow$ $\begin{pmatrix} 0 & -2\gamma+\triangle \\ 2\gamma+\triangle & 0 \end{pmatrix}\\$ $\Rightarrow$ $\begin{pmatrix} 0 & 2\gamma+\triangle \\ -2\gamma+\triangle & 0 \end{pmatrix}\\$
where $\triangle = 2t_{1}\cos(k)+2t_{2}\cos(2k)	$.

\begin{figure}[!htbp]
\includegraphics[width=4.2cm,height=3.8cm]{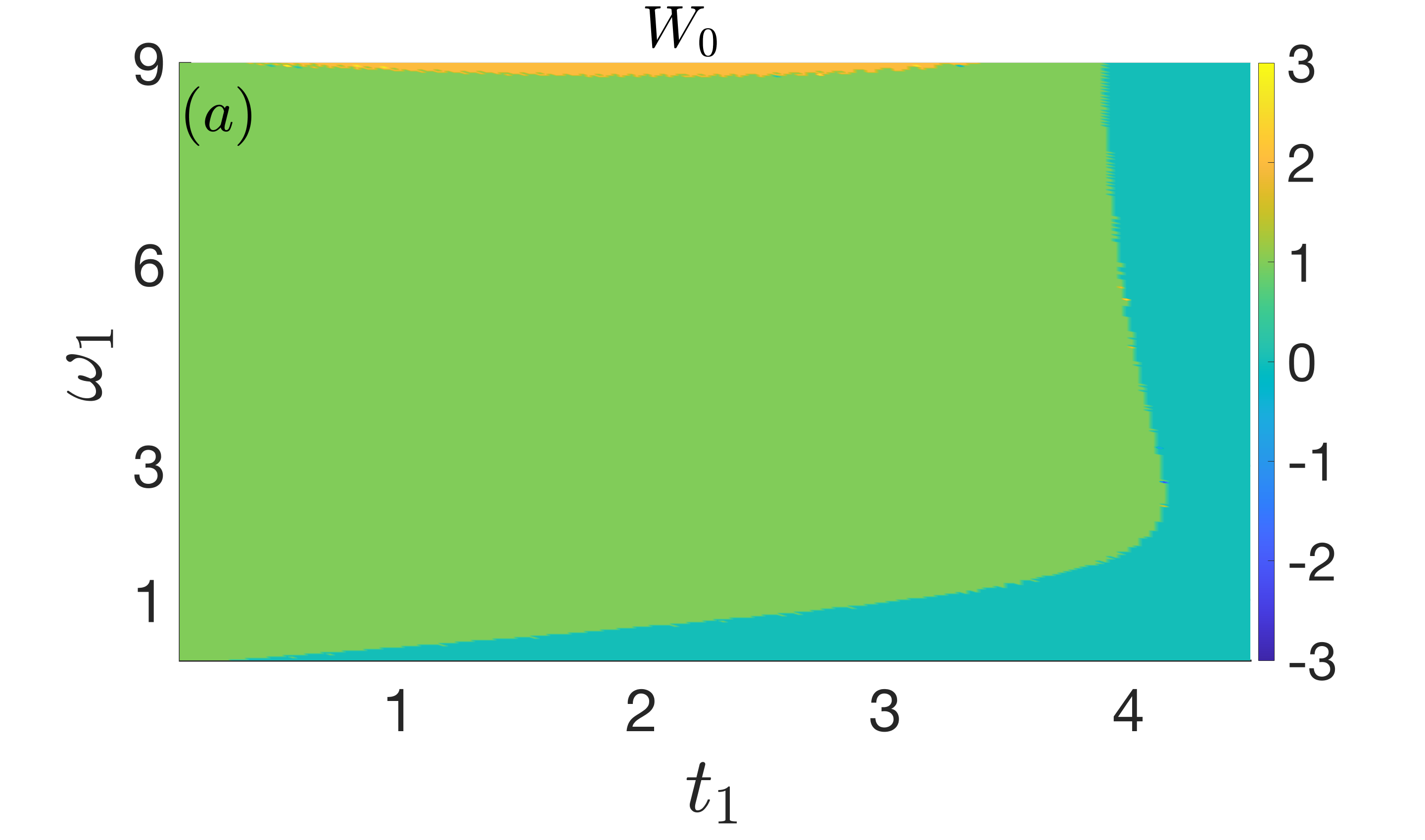}
\includegraphics[width=4.2cm,height=3.8cm]{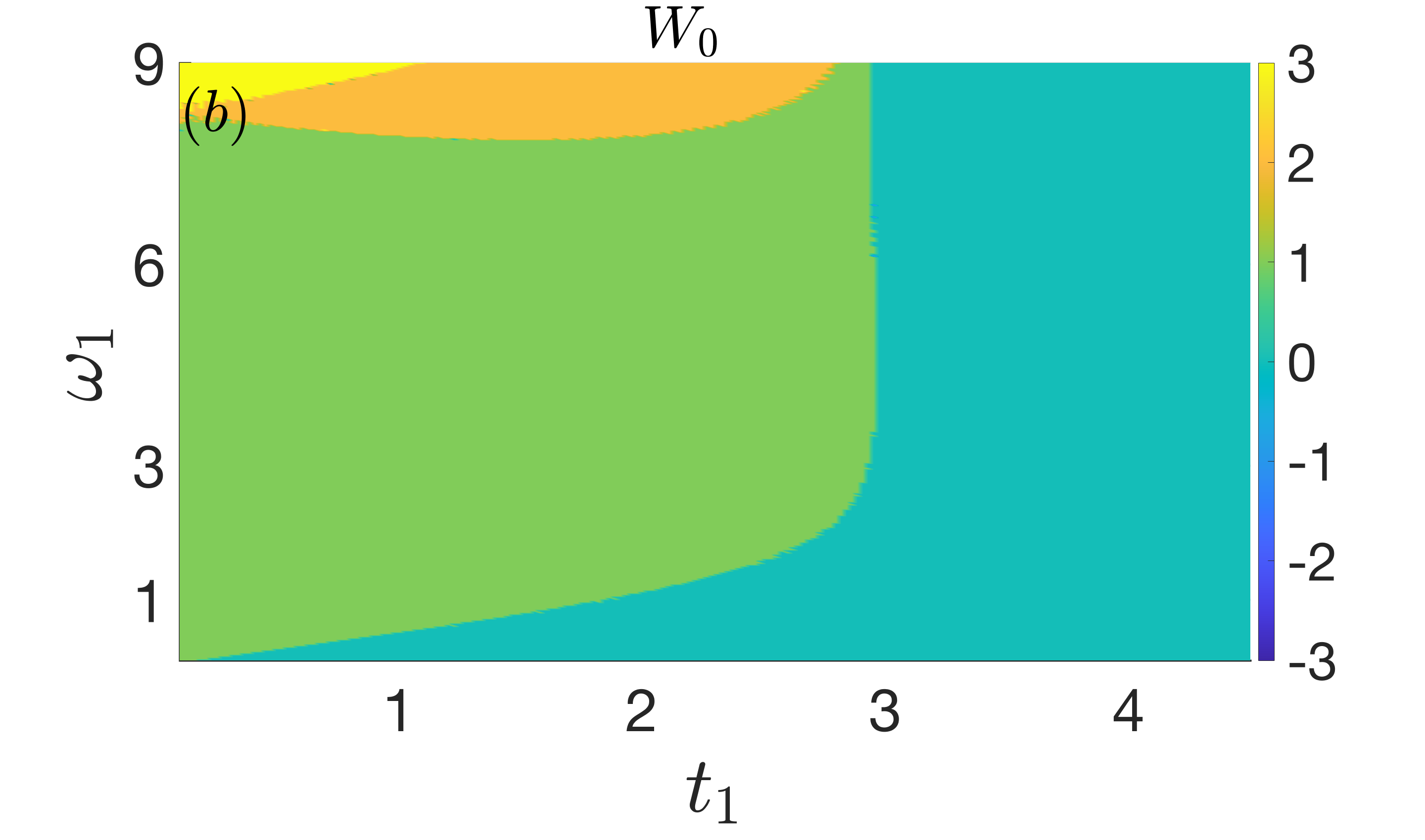}
\includegraphics[width=4.2cm,height=3.8cm]{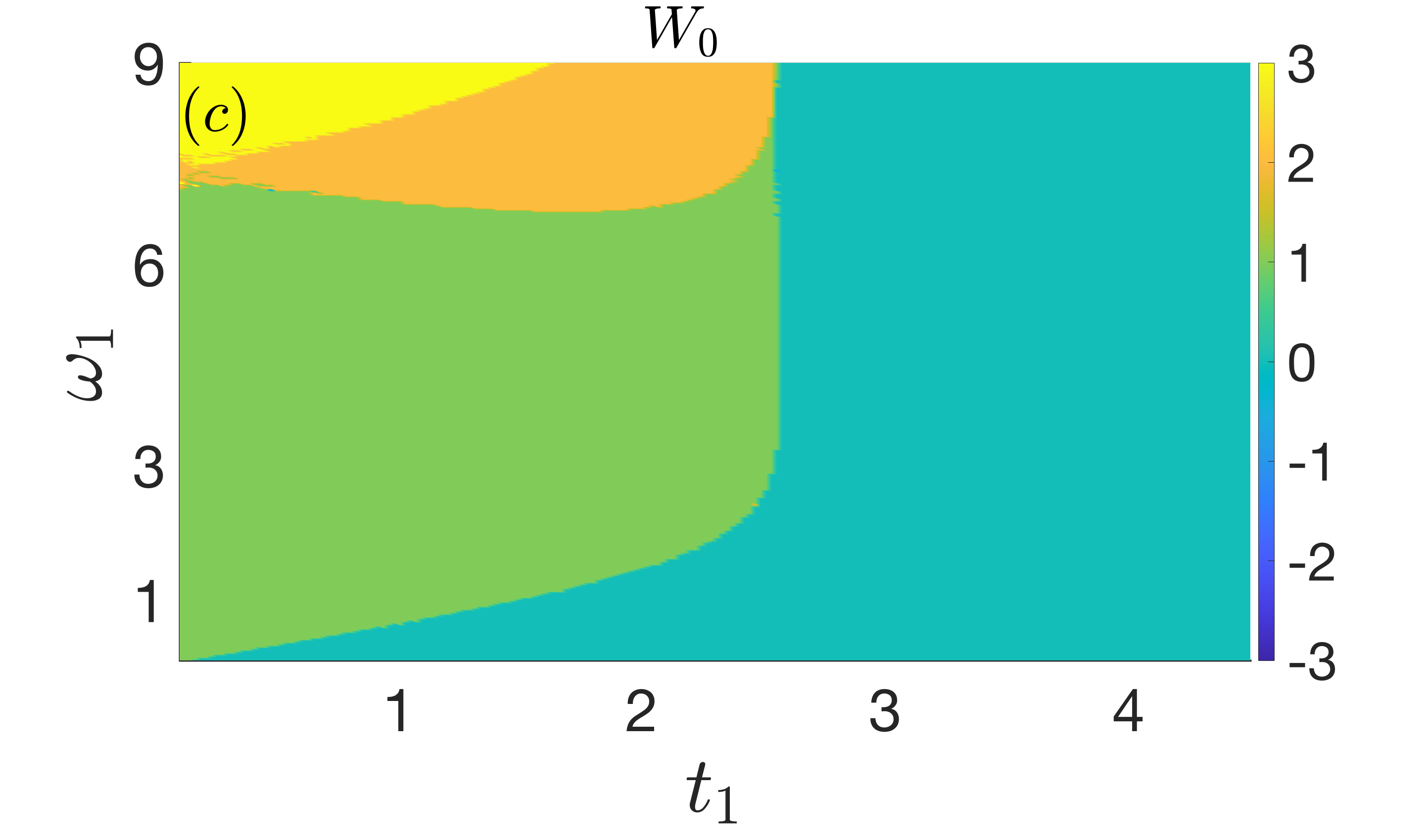}

\includegraphics[width=4.2cm,height=3.8cm]{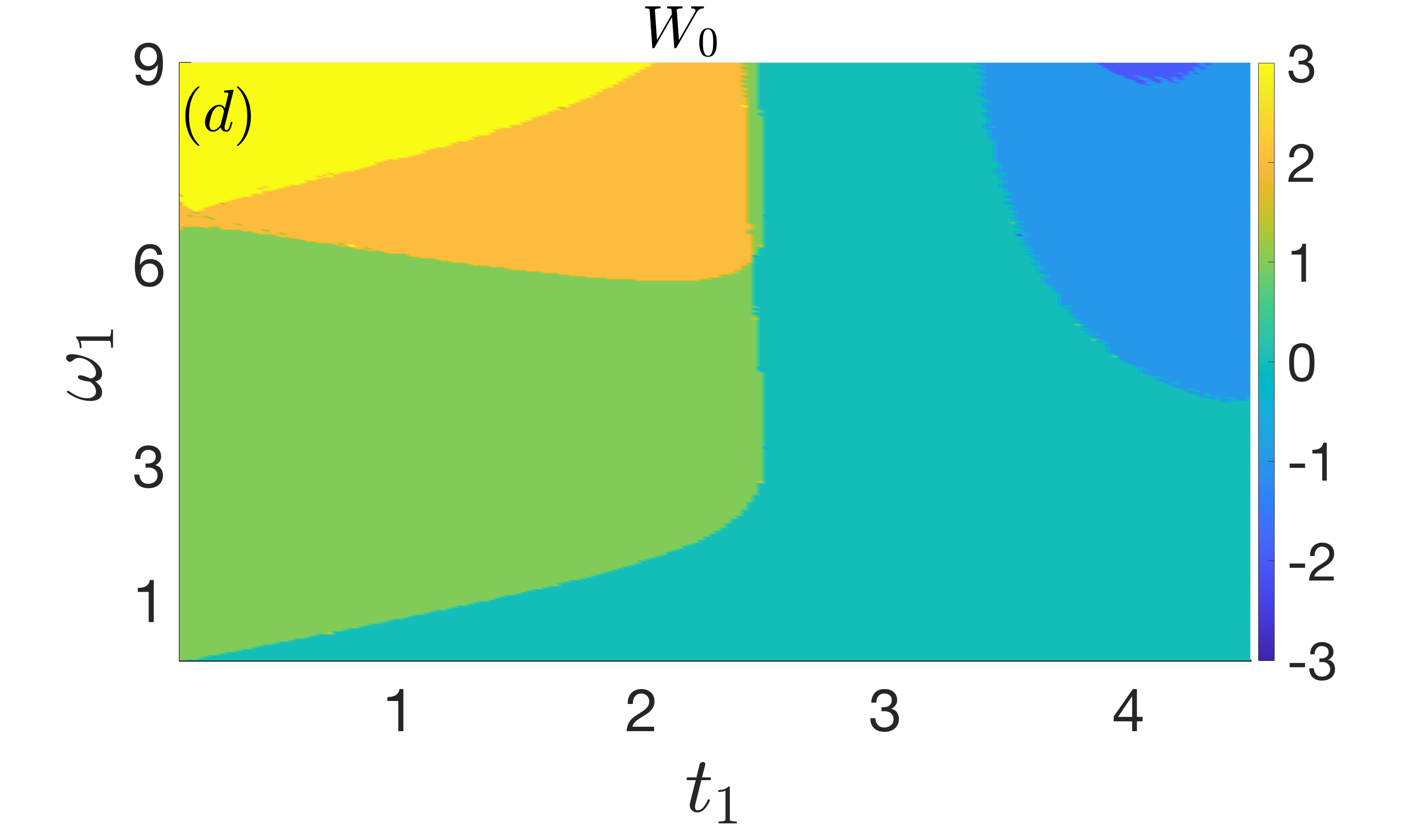}
\includegraphics[width=4.2cm,height=3.8cm]{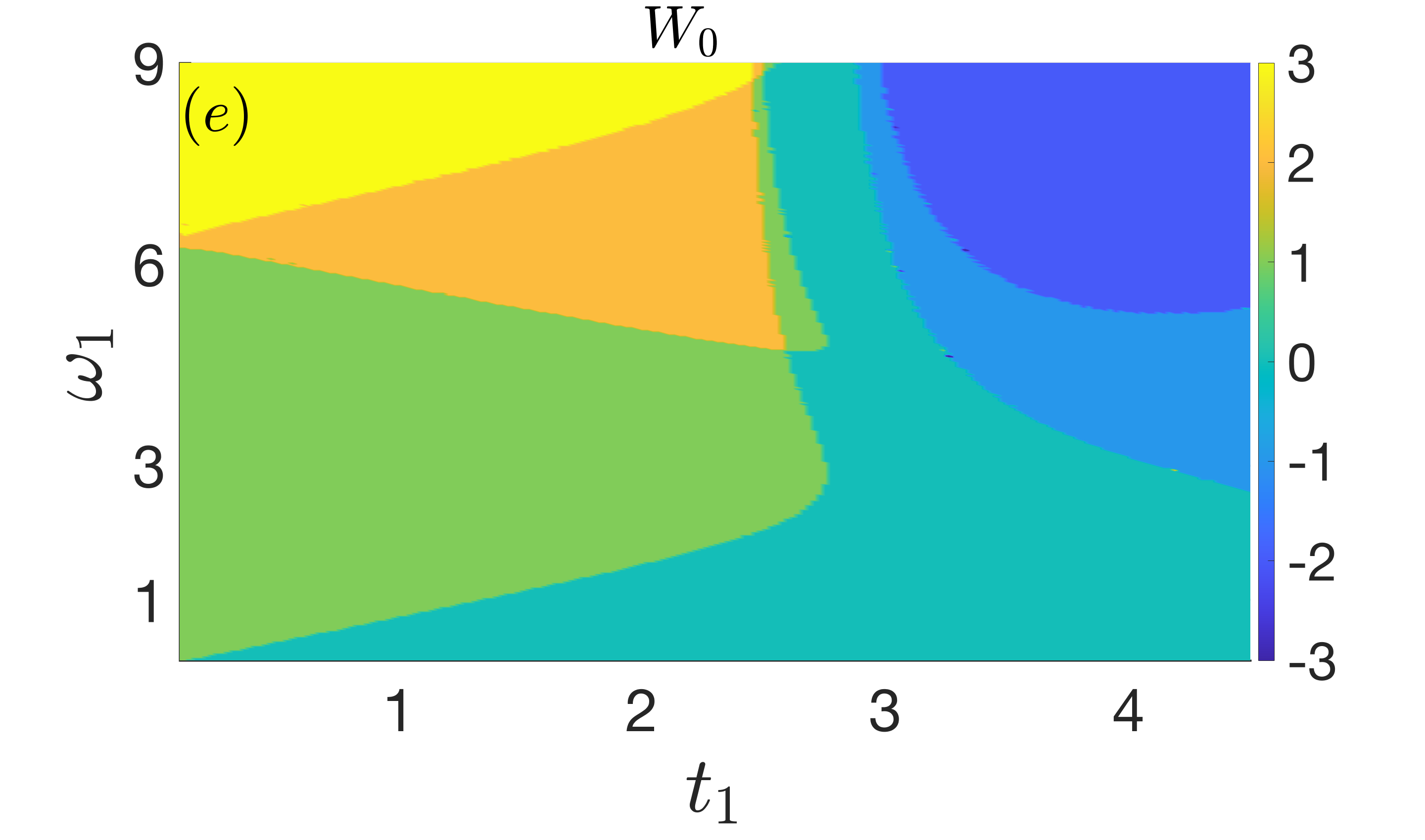}
\includegraphics[width=4.2cm,height=3.8cm]{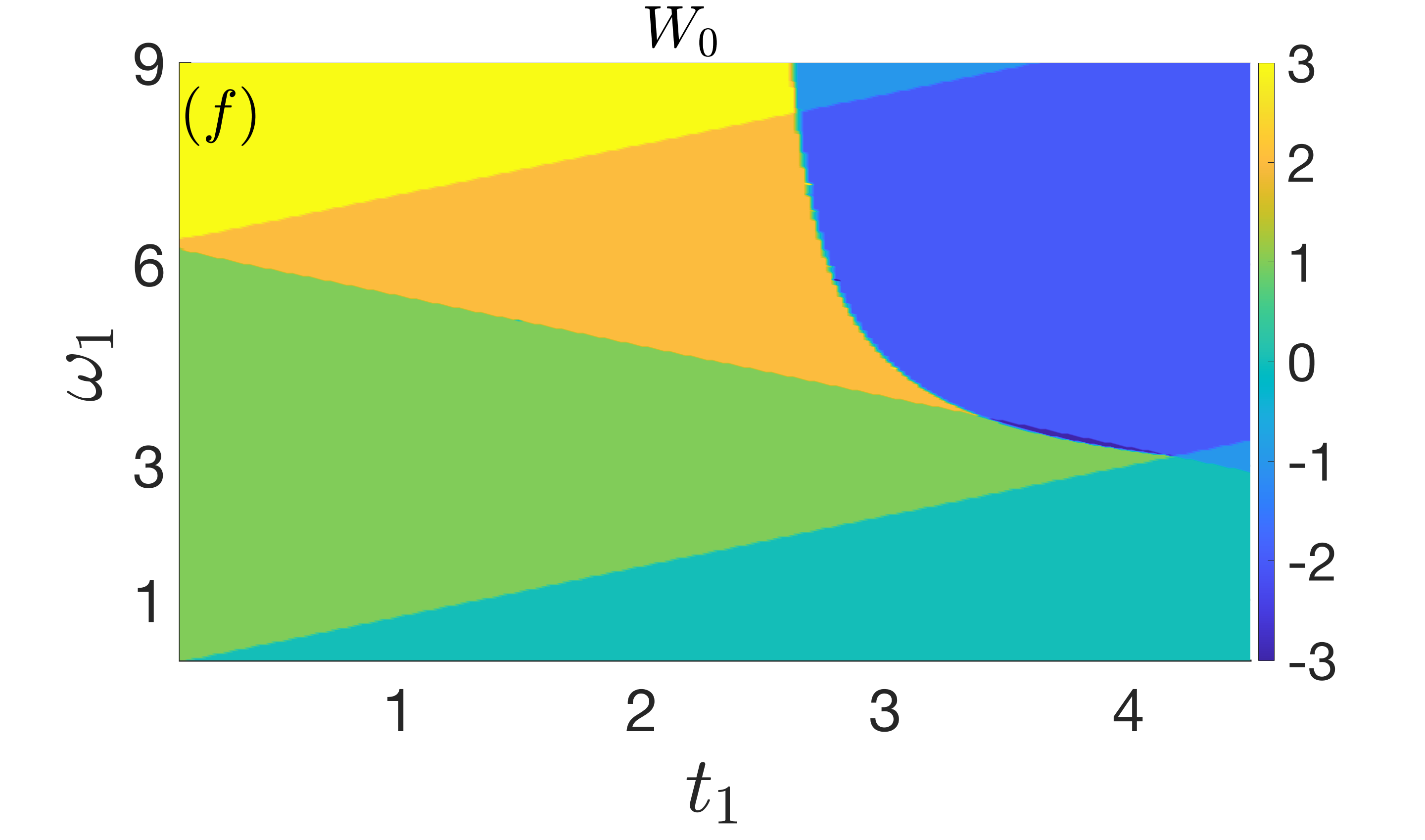}
\caption{(Color online) A pair of winding numbers $W_{0}$ versus hopping amplitudes $t_{1}$ and $\omega_{1}$, the common parameters are given by $N=200$, $t_{2}=0.01$, $\mu=0$, $\omega_{2}=0.01$. Intracell asymmetric hopping amplitude $\gamma=\gamma_{0}\cos{\theta}+i\gamma_{0}\sin{\theta}$, where $\gamma_{0}=0.75$ and the phase $\theta=\frac{\pi}{12}, \frac{\pi}{6}, \frac{\pi}{4}, \frac{\pi}{3}, \frac{5\pi}{12}, \frac{\pi}{2}$.}
\label{fig3}
\end{figure}

Though the purely imaginary asymmetric term $\gamma$ in Hamilton does not cause Non-Hermiticity, the introduction of the purely imaginary hopping term induced by a phase $\frac{\pi}{2}$ can bring an extremely rich phase diagram in Figs. \ref{fig2} (c)-(d), which is in sharp contrast with the above two graphs. The values of $W_{0}$ and $W_{\pi}$ are no longer as monotonous as the ones of the real hopping term. This system generates phases with larger topological numbers, as $t_{1}$ takes a value from 0 to 4 and $\omega_{1}$ takes a value from 0 to 9, $W_{0}$ and $W_{\pi}$ takes all integers from -3 to 3. Numerical results show that when the values of $t_{1}$  and $\omega_{1}$ are not restricted, Floquet topological phases induced by a purely imaginary number will hold unlimited winding numbers.

Next, we will discuss the situation when the complex parameter $\gamma$ is no longer a pure imaginary number, that is, the real part is not 0. According to the symmetry of the trigonometric function, we only consider the value of $\theta$ in the interval 0 to $\frac{\pi}{2}$. Figs. \ref{fig3} (a)-(f) show $W_{0}$ under different phases as $\theta$ increases from 0 to $\frac{\pi}{2}$. As the modulus of $\gamma$ does not change, the topological phase transitions in this case are due to the non-trivial phase $\theta$. Here, the Non-Hermitianity is still caused by the real part of $\gamma$. Comparing Fig. \ref{fig3} with Fig. \ref{fig2}(a), one can find that phases with $W_{0}=\pm2,\pm3$ appear. The system has a tendency to produce phases with larger topological numbers, and phases with small topological numbers will transform to phases with a larger topological number caused by the non-trivial phase. The similar phenomenon of $W_{\pi}$  and the unique sensitivity to phase changes within a certain range is present in Appendix \ref{Appendix A}.

\begin{figure}[!htbp]
\includegraphics[width=4.2cm,height=3.8cm]{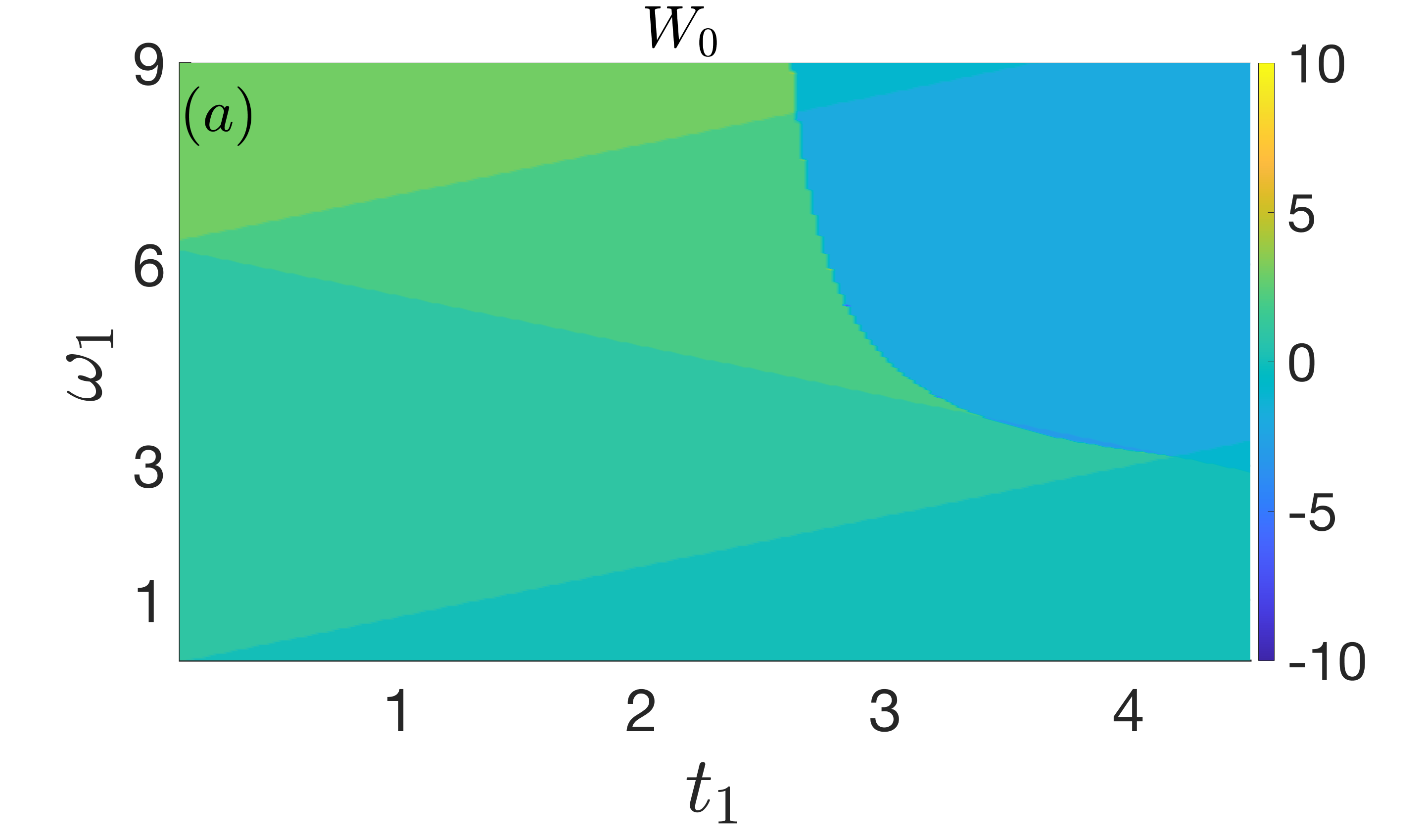}
\includegraphics[width=4.2cm,height=3.8cm]{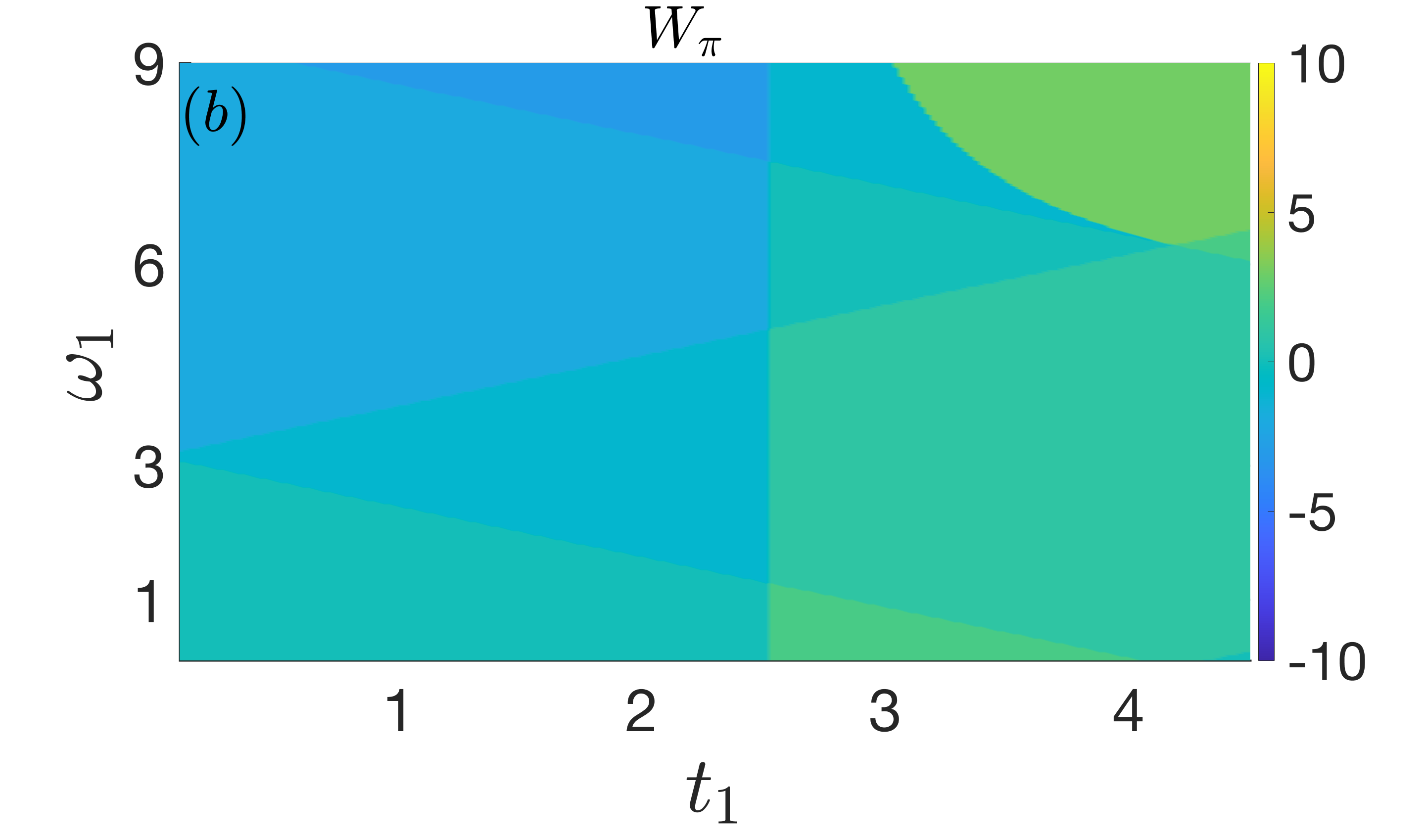}
\includegraphics[width=4.2cm,height=3.8cm]{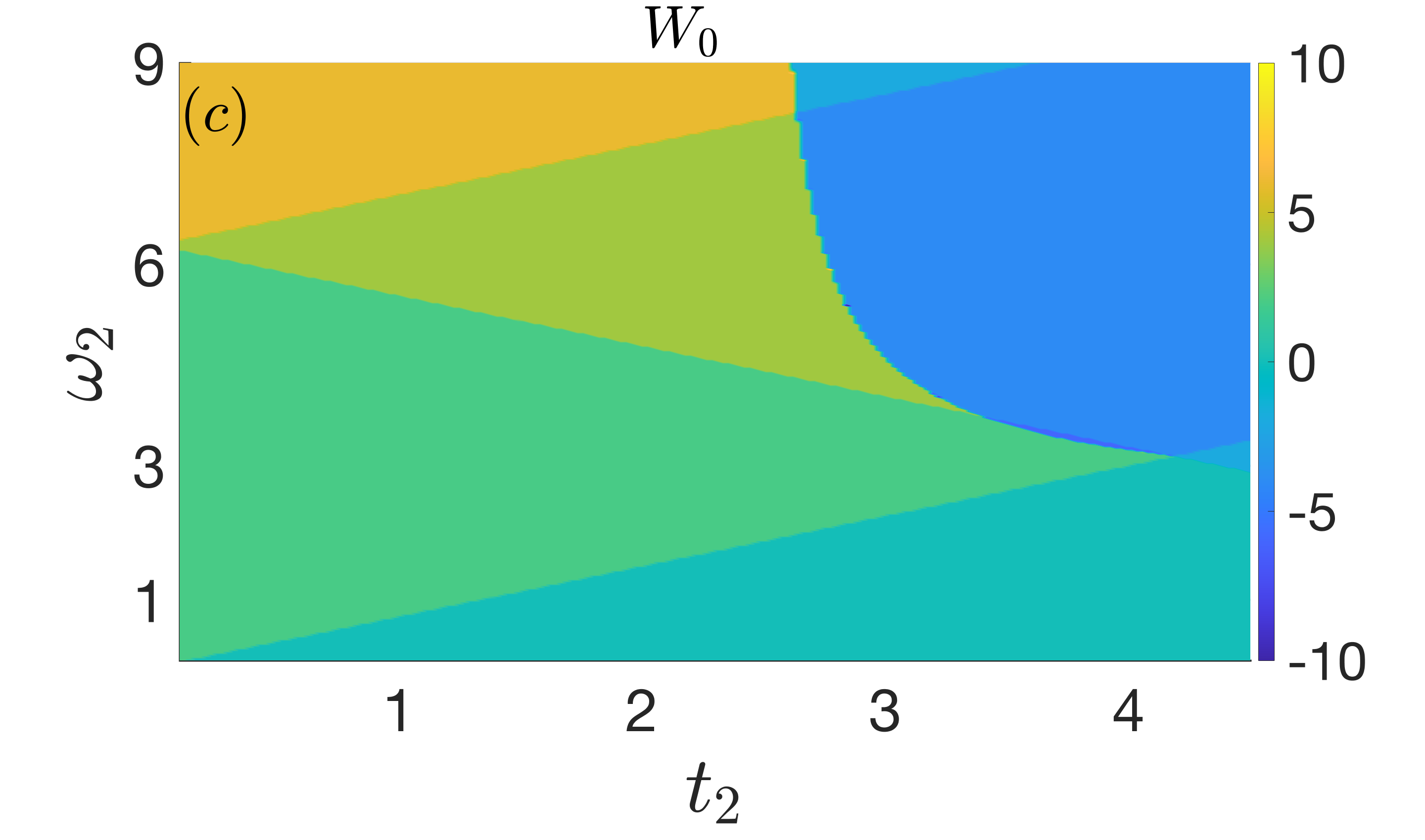}
\includegraphics[width=4.2cm,height=3.8cm]{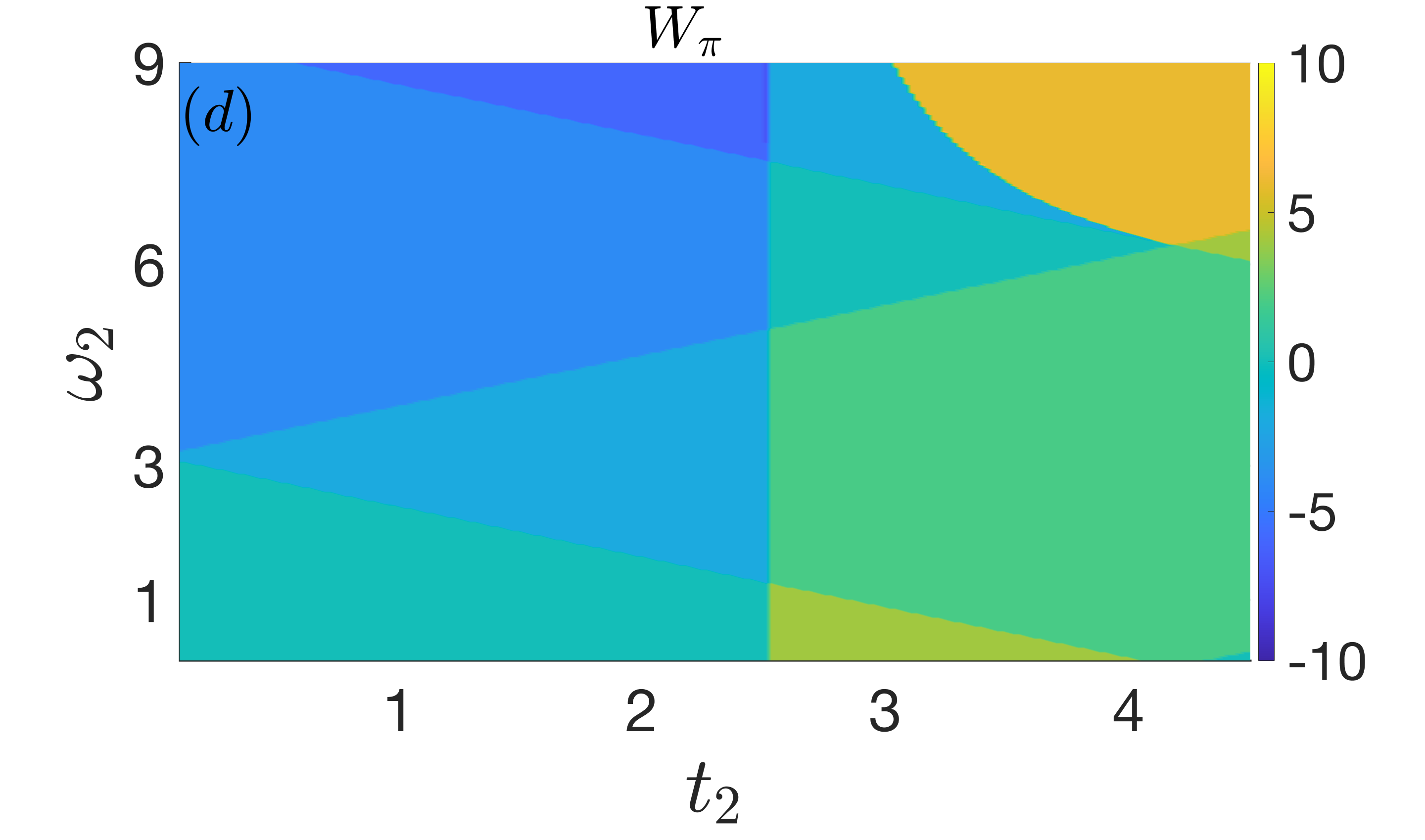}

\caption{(Color online) For (a)-(b). A pair of winding numbers with system hopping amplitudes $t_{1}$ and $\omega_{1}$, the common parameters are given by $N=200$, $t_{2}=0$, $\mu=0$, $\omega_{2}=0$ and $\gamma = 0.75i*t_{1}$. For (c)-(d). A pair of winding numbers as functions of hopping amplitudes $t_{2}$ and $\omega_{2}$, the common parameters are given by $N=200$, $t_{1}=0$, $\mu=0$, $\omega_{1}=0$ and $\gamma = 0.75i*t_{2}$.}
\label{fig4}
\end{figure}

Further, the topological properties of the Non-Hermitian Floquet system with long-range hopping amplitudes inspire much interest, which will be explored in this part. For comparison, we first only take the nearest-neighbor hopping amplitudes into account for simplicity, that is, $t_{2}=0$ and $\omega_{2}=0$. As shown in Figs. \ref{fig4}(a)-(b), we plot a pair of winding numbers as functions of the hopping amplitudes $t_{2}$ and $\omega_{1}$.  In Figs. \ref{fig4}(c)-(d), we present a pair of winding numbers with only next-nearest hopping amplitudes. Obviously, the value of the winding number is twice that of considering only the nearest neighbor hopping amplitudes, that is, winding number $W_{0}(2)=2W_{0}$, $W_{\pi}(2)=2W_{\pi}$ where $W_{0/\pi}(2)$ refers to winding number containing only the next-nearest hopping term. As we mentioned above, if the parameters of the system are not restricted, then we will get an unlimited topological winding numbers. When we only consider the next-nearest neighbor hopping term, one can get even winding numbers and there is no odd winding numbers. This effect of selecting even topological numbers is very practical to meet certain requirements.

\section{Conclusion and discussion}\label{V}

In summary, we have explored a Non-Hermitian long-range system with complex hopping amplitudes, which is subjected to piecewise time-periodic quenches. We have revealed that the winding number $W_{0}$ and $W_{\pi}$ calculated using the Bloch band theory are equal to the number of edge state pairs at quasienergies zero and $\pi$ even though the Non-Hermitian is considered. Meanwhile, the introduction of pure imaginary terms can degenerate the system into Hermitian, and bring extremely rich Floquet topological phases. Furthermore, the system can generate a topological phase with a large topological number when the value of the phase increases. Moreover, the next-nearest hopping term plays a role in selecting an even topological number from unlimited winding numbers.

Although the Hamiltonian \eqref{1} is simple here, the results reflected should be thought-provoking. Such as, one can ever get richer topological phase just changing the phase of the hopping simply, and this phenomenon may have a great impact on the transport properties of the system [\onlinecite{Gro74}]. In addition, the combination of the long-range hopping and the Floquet engineering endows the system the function of selecting the even topological numbers. Then, an open question arises: How to only obtain the odd topological numbers? More generally, How to obtain the topological numbers arranged in a form of the arithmetic progression with tolerances $d$? Our findings are the first step towards the understanding of such matters.

\section{ACKNOWLEDGMENTS}\label{VI}

This work was supported by National Natural Science Foundation of China (Grants No. 11874190, No. 61835013 and No. 12047501), and National Key R\&D Program of China under grants No. 2016YFA0301500. Support was also provided by Supercomputing Center of Lanzhou University.

\appendix
\section{Exotic behavior of $W_{\pi}$}\label{Appendix A}

\begin{figure}[!htbp]
\includegraphics[width=4.2cm,height=3.8cm]{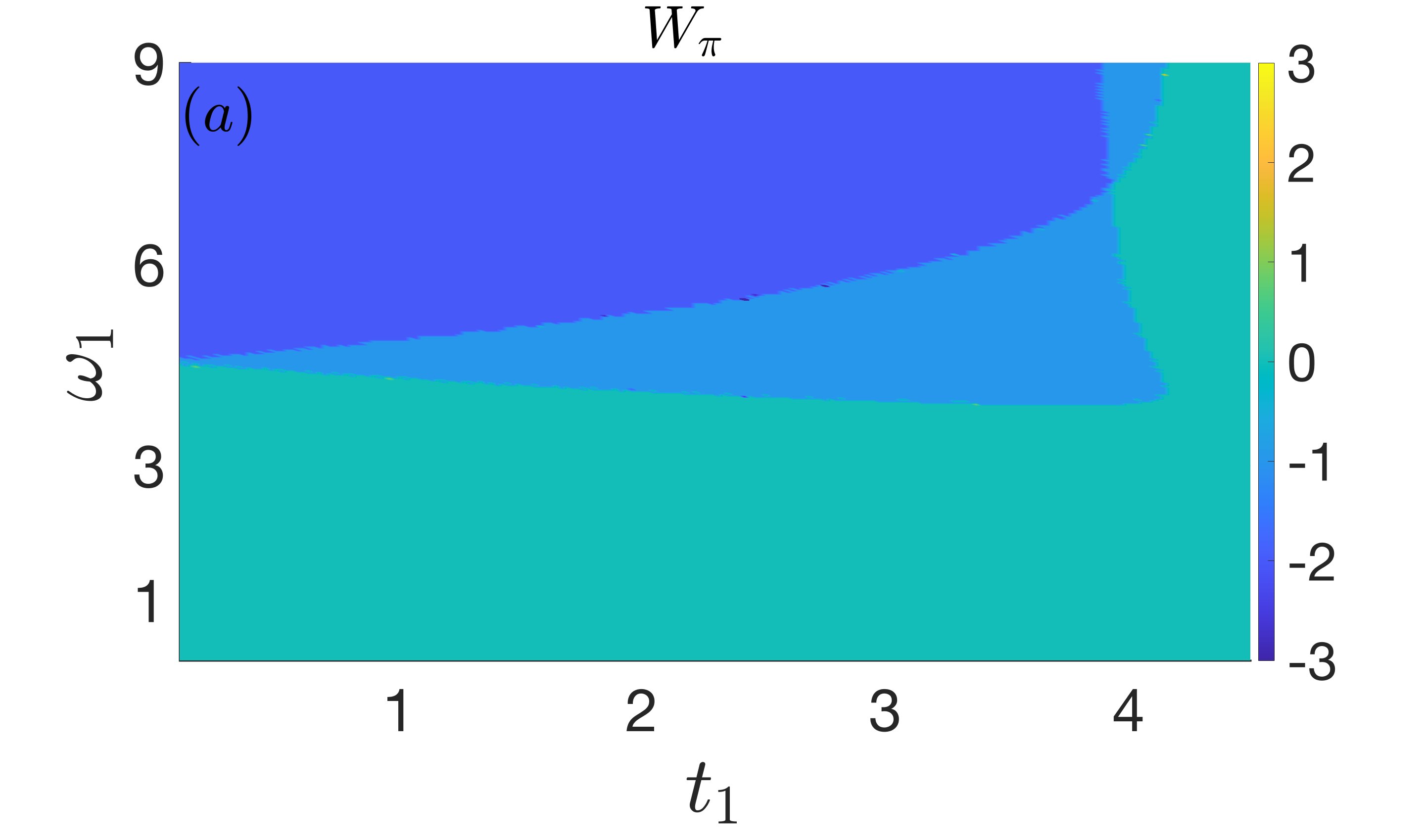}
\includegraphics[width=4.2cm,height=3.8cm]{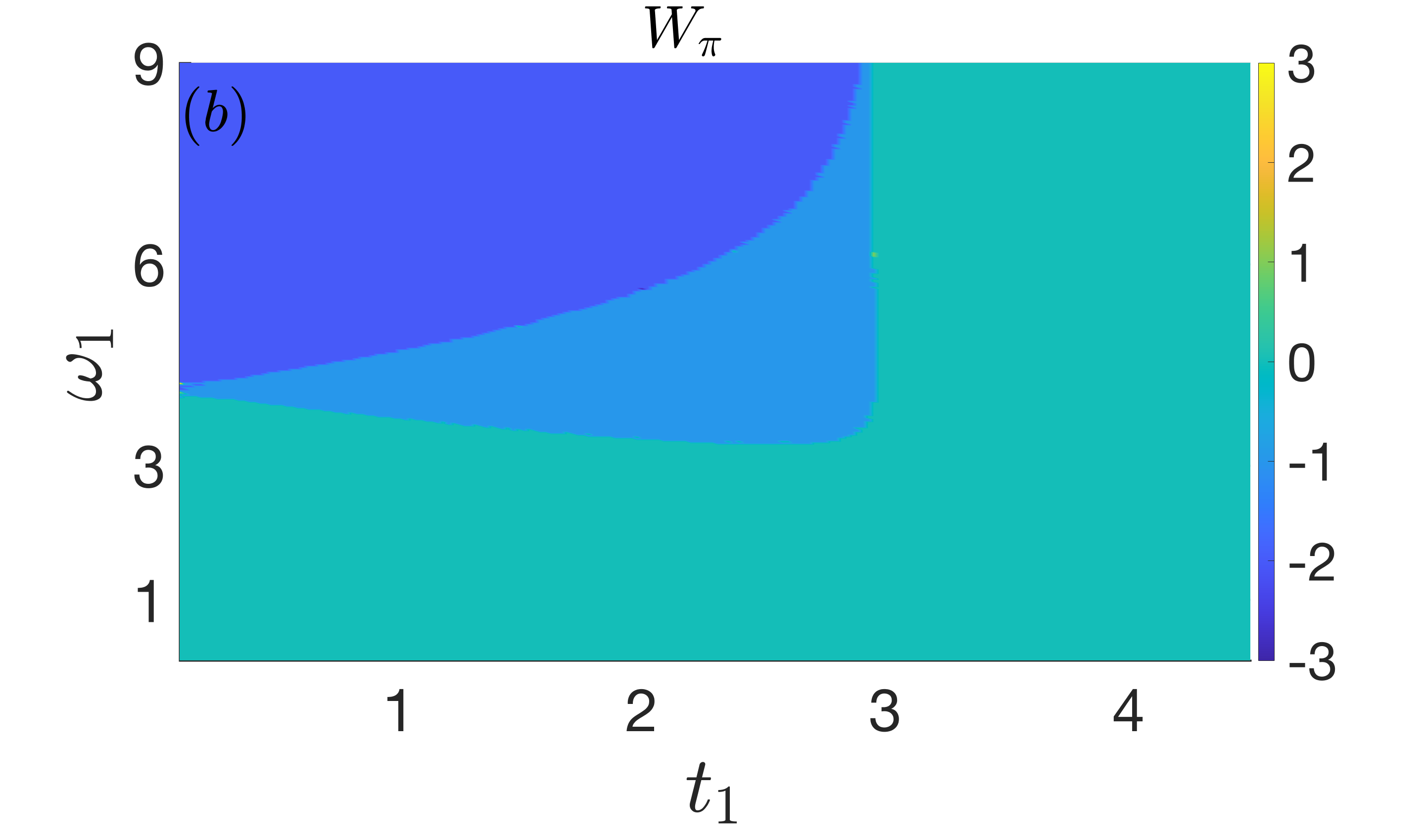}
\includegraphics[width=4.2cm,height=3.8cm]{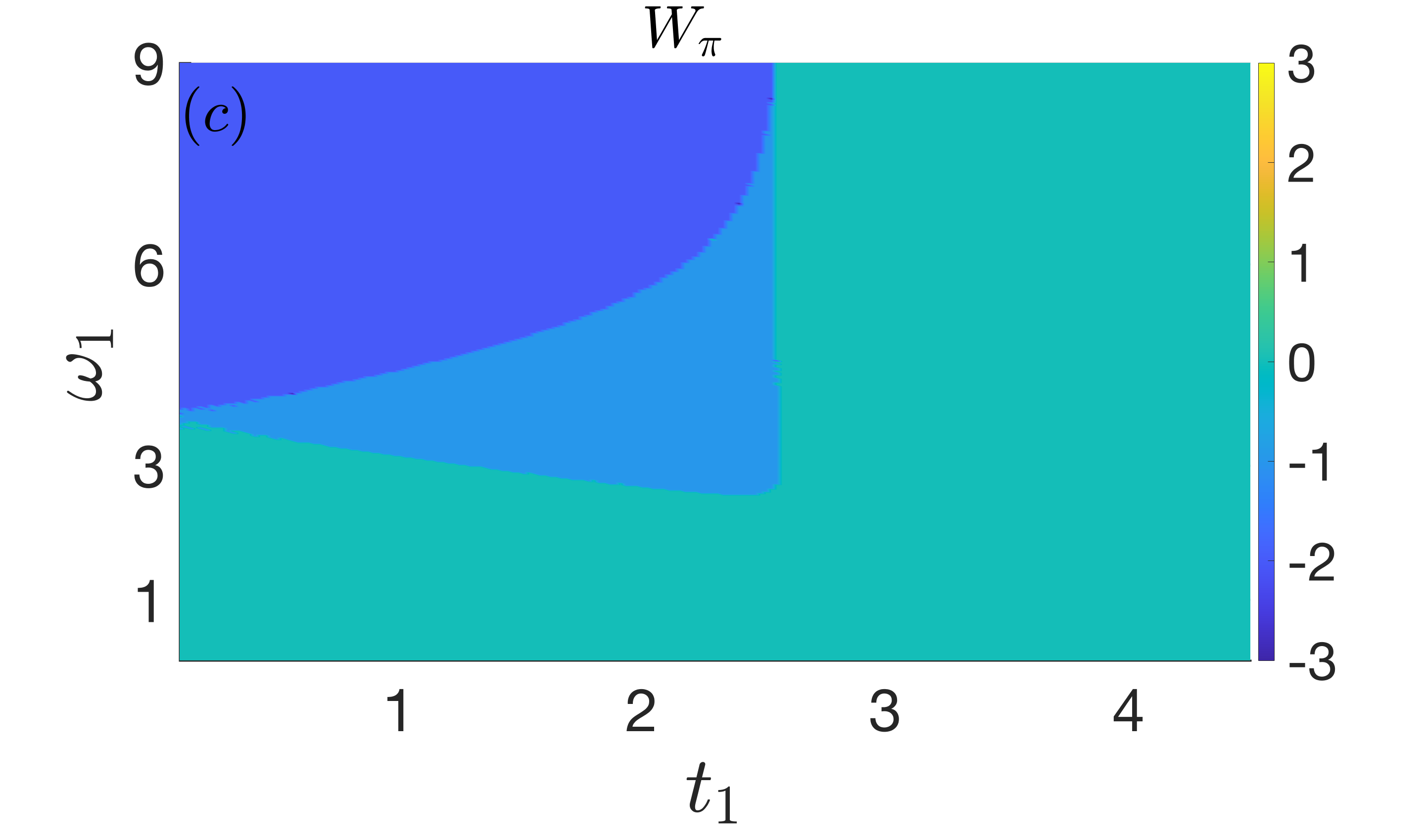}

\includegraphics[width=4.2cm,height=3.8cm]{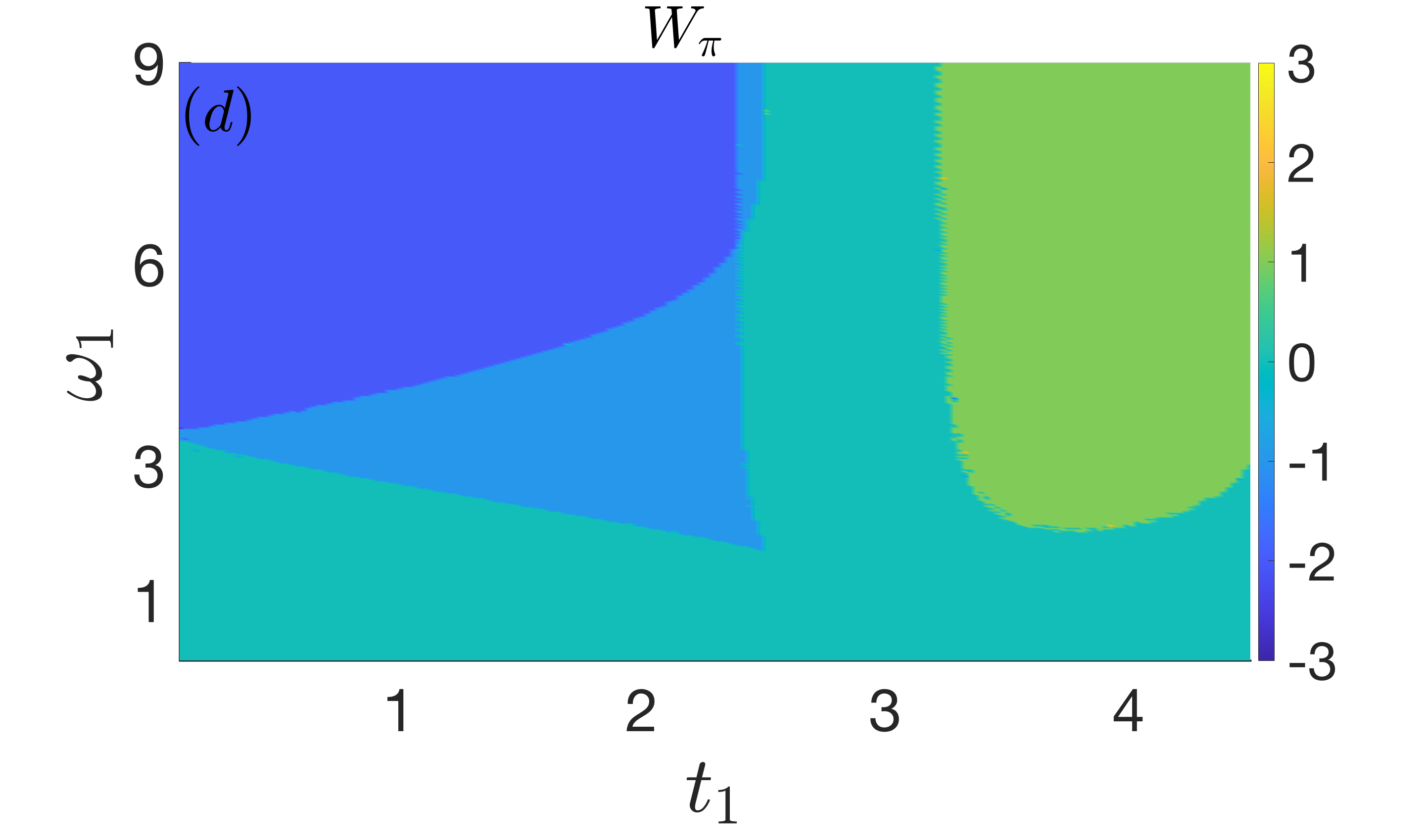}
\includegraphics[width=4.2cm,height=3.8cm]{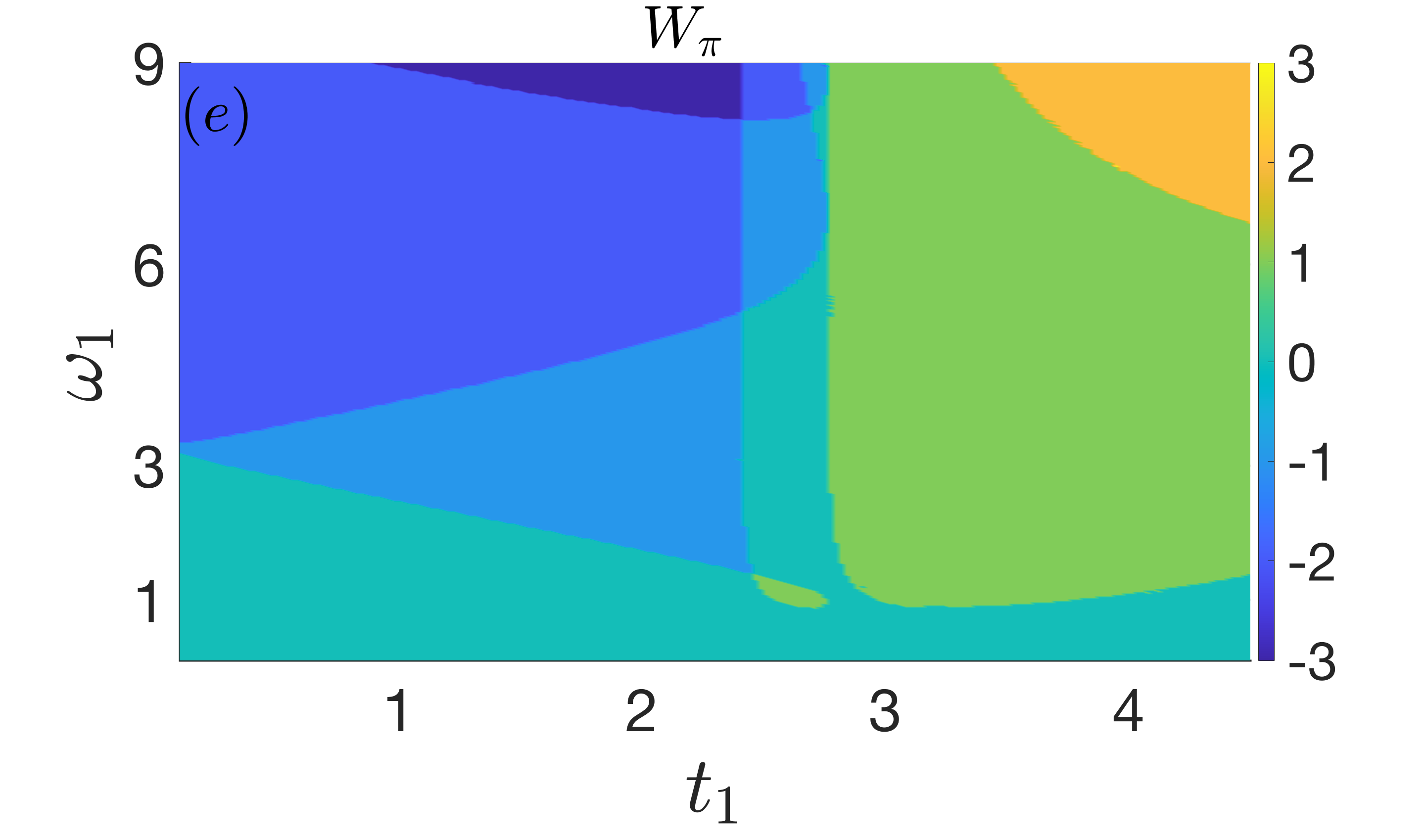}
\includegraphics[width=4.2cm,height=3.8cm]{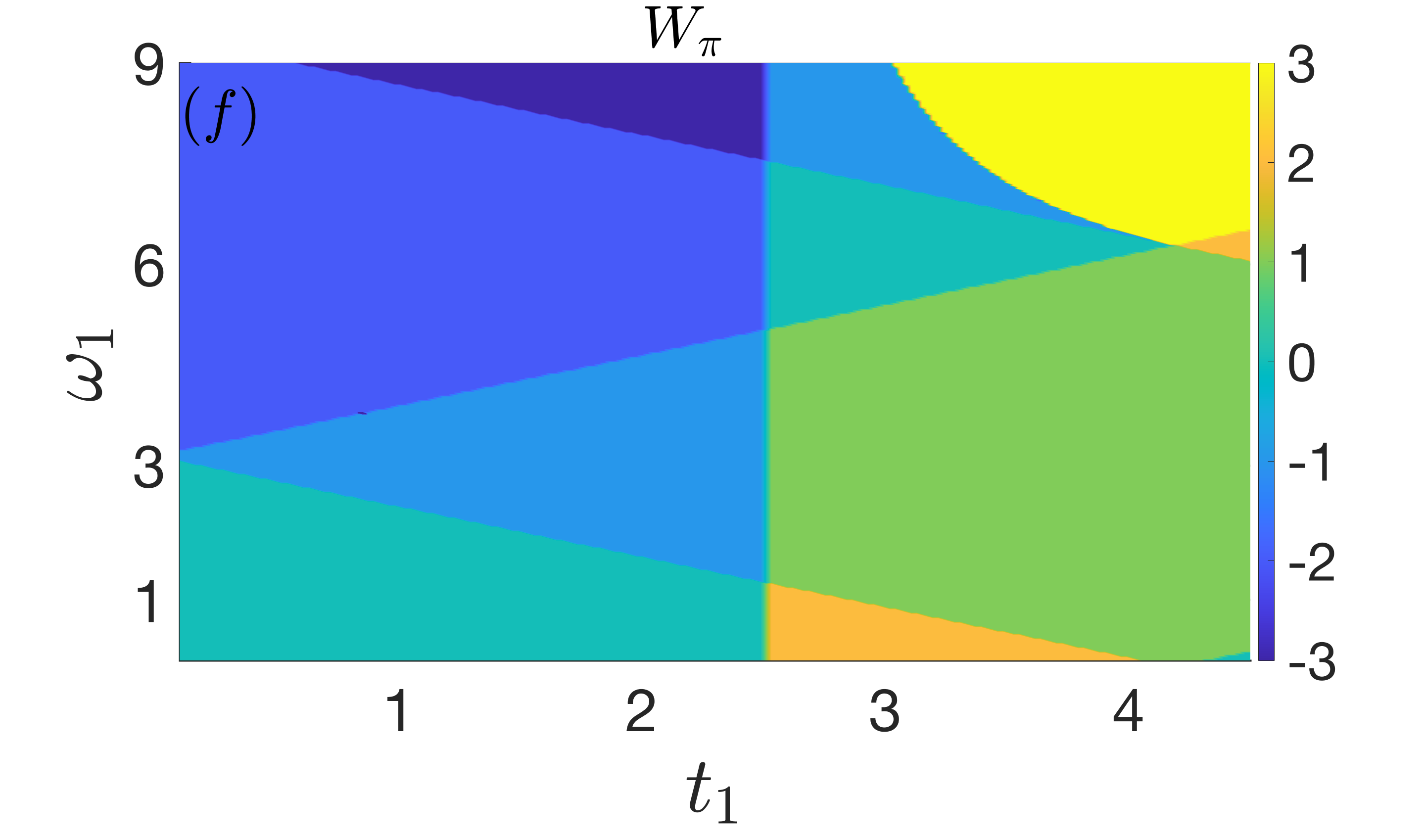}
\caption{(Color online) A pair of winding numbers $W_{\pi}$ with system hopping amplitudes $t_{1}$ and $\omega_{1}$, the common parameters are given by $N=200$, $t_{2}=0.01$, $\mu=0$, $\omega_{2}=0.01$. Intracell asymmetric hopping amplitude $\gamma=\gamma_{0}\cos{\theta}+i\gamma_{0}\sin{\theta}$, where $\gamma_{0}=0.75$ and the phase $\theta=\frac{\pi}{12}, \frac{\pi}{6}, \frac{\pi}{4}, \frac{\pi}{3}, \frac{5\pi}{12}, \frac{\pi}{2}$.}
\label{fig5}
\end{figure}

We show the winding number $W_{\pi}$ versus the system hopping amplitudes $t_{1}$ and $\omega_{1}$ in Figs. \ref{fig5}(a)-(f). We can find that in the same phase change interval $(0, 2\pi]$,  $W_{\pi}$  exhibits similar properties to  $W_{0}$. As the phase increases,  $W_{\pi}$ also has a tendency to transform to a phase with a large topological number. However, $W_{\pi}$ exhibits unique sensitivity to the changes in phase values within a certain range. As the phase $\theta$ increases, the phase areas of $W_{\pi}=1,2$ in the phase diagram slowly decrease, and $W_{\pi}$ does not transform toward a phase with a large topological number. When the phase value increases to close to $\frac{\pi}{3}$, $W_{\pi}$ Rapidly adds new topological phases, and these topological phases have a larger topological numbers like $\pm3$. We can see that $W_{\pi}$ still has a tendency to produce topological phases with larger topological numbers, and the phase of the hopping can cause topological phase transitions.

\bibliography{refFl}

\begin{thebibliography}{83}%
\makeatletter
\providecommand \@ifxundefined [1]{%
 \@ifx{#1\undefined}
}%
\providecommand \@ifnum [1]{%
 \ifnum #1\expandafter \@firstoftwo
 \else \expandafter \@secondoftwo
 \fi
}%
\providecommand \@ifx [1]{%
 \ifx #1\expandafter \@firstoftwo
 \else \expandafter \@secondoftwo
 \fi
}%
\providecommand \natexlab [1]{#1}%
\providecommand \enquote  [1]{``#1''}%
\providecommand \bibnamefont  [1]{#1}%
\providecommand \bibfnamefont [1]{#1}%
\providecommand \citenamefont [1]{#1}%
\providecommand \href@noop [0]{\@secondoftwo}%
\providecommand \href [0]{\begingroup \@sanitize@url \@href}%
\providecommand \@href[1]{\@@startlink{#1}\@@href}%
\providecommand \@@href[1]{\endgroup#1\@@endlink}%
\providecommand \@sanitize@url [0]{\catcode `\\12\catcode `\$12\catcode
  `\&12\catcode `\#12\catcode `\^12\catcode `\_12\catcode `\%12\relax}%
\providecommand \@@startlink[1]{}%
\providecommand \@@endlink[0]{}%
\providecommand \url  [0]{\begingroup\@sanitize@url \@url }%
\providecommand \@url [1]{\endgroup\@href {#1}{\urlprefix }}%
\providecommand \urlprefix  [0]{URL }%
\providecommand \Eprint [0]{\href }%
\providecommand \doibase [0]{http://dx.doi.org/}%
\providecommand \selectlanguage [0]{\@gobble}%
\providecommand \bibinfo  [0]{\@secondoftwo}%
\providecommand \bibfield  [0]{\@secondoftwo}%
\providecommand \translation [1]{[#1]}%
\providecommand \BibitemOpen [0]{}%
\providecommand \bibitemStop [0]{}%
\providecommand \bibitemNoStop [0]{.\EOS\space}%
\providecommand \EOS [0]{\spacefactor3000\relax}%
\providecommand \BibitemShut  [1]{\csname bibitem#1\endcsname}%
\let\auto@bib@innerbib\@empty
\bibitem [{\citenamefont {{P. A. M. Dirac}}(1981)}]{1}%
  \BibitemOpen
  \bibfield  {author} {\bibinfo {author} {\bibnamefont {{P. A. M. Dirac}}},\
  }\href@noop {} {\emph {\bibinfo {title} {{The Principles of Quantum
  Mechanics}}}}\ (\bibinfo  {publisher} {{Clarendon, Oxford}},\ \bibinfo {year}
  {1981})\BibitemShut {NoStop}%
\bibitem [{\citenamefont {Agarwal}\ and\ \citenamefont {Joglekar}(2021)}]{11}%
  \BibitemOpen
  \bibfield  {author} {\bibinfo {author} {\bibfnamefont {K.~S.}\ \bibnamefont
  {Agarwal}}\ and\ \bibinfo {author} {\bibfnamefont {Y.~N.}\ \bibnamefont
  {Joglekar}},\ }\href {\doibase 10.1103/PhysRevA.104.022218} {\bibfield
  {journal} {\bibinfo  {journal} {Phys. Rev. A}\ }\textbf {\bibinfo {volume}
  {104}},\ \bibinfo {pages} {022218} (\bibinfo {year} {2021})}\BibitemShut
  {NoStop}%
\bibitem [{\citenamefont {Zeng}\ \emph {et~al.}(2017)\citenamefont {Zeng},
  \citenamefont {Chen},\ and\ \citenamefont {L\"u}}]{12}%
  \BibitemOpen
  \bibfield  {author} {\bibinfo {author} {\bibfnamefont {Q.-B.}\ \bibnamefont
  {Zeng}}, \bibinfo {author} {\bibfnamefont {S.}~\bibnamefont {Chen}}, \ and\
  \bibinfo {author} {\bibfnamefont {R.}~\bibnamefont {L\"u}},\ }\href {\doibase
  10.1103/PhysRevA.95.062118} {\bibfield  {journal} {\bibinfo  {journal} {Phys.
  Rev. A}\ }\textbf {\bibinfo {volume} {95}},\ \bibinfo {pages} {062118}
  (\bibinfo {year} {2017})}\BibitemShut {NoStop}%
\bibitem [{\citenamefont {Wang}\ \emph {et~al.}(2018)\citenamefont {Wang},
  \citenamefont {Ye}, \citenamefont {Christensen},\ and\ \citenamefont
  {Liu}}]{13}%
  \BibitemOpen
  \bibfield  {author} {\bibinfo {author} {\bibfnamefont {M.}~\bibnamefont
  {Wang}}, \bibinfo {author} {\bibfnamefont {L.}~\bibnamefont {Ye}}, \bibinfo
  {author} {\bibfnamefont {J.}~\bibnamefont {Christensen}}, \ and\ \bibinfo
  {author} {\bibfnamefont {Z.}~\bibnamefont {Liu}},\ }\href {\doibase
  10.1103/PhysRevLett.120.246601} {\bibfield  {journal} {\bibinfo  {journal}
  {Phys. Rev. Lett.}\ }\textbf {\bibinfo {volume} {120}},\ \bibinfo {pages}
  {246601} (\bibinfo {year} {2018})}\BibitemShut {NoStop}%
\bibitem [{\citenamefont {Longhi}(2013)}]{14}%
  \BibitemOpen
  \bibfield  {author} {\bibinfo {author} {\bibfnamefont {S.}~\bibnamefont
  {Longhi}},\ }\href {\doibase 10.1103/PhysRevA.88.052102} {\bibfield
  {journal} {\bibinfo  {journal} {Phys. Rev. A}\ }\textbf {\bibinfo {volume}
  {88}},\ \bibinfo {pages} {052102} (\bibinfo {year} {2013})}\BibitemShut
  {NoStop}%
\bibitem [{\citenamefont {Gu}\ \emph {et~al.}(2021)\citenamefont {Gu},
  \citenamefont {Gao}, \citenamefont {Cao}, \citenamefont {Liu}, \citenamefont
  {Zhu},\ and\ \citenamefont {Zhu}}]{15}%
  \BibitemOpen
  \bibfield  {author} {\bibinfo {author} {\bibfnamefont {Z.}~\bibnamefont
  {Gu}}, \bibinfo {author} {\bibfnamefont {H.}~\bibnamefont {Gao}}, \bibinfo
  {author} {\bibfnamefont {P.-C.}\ \bibnamefont {Cao}}, \bibinfo {author}
  {\bibfnamefont {T.}~\bibnamefont {Liu}}, \bibinfo {author} {\bibfnamefont
  {X.-F.}\ \bibnamefont {Zhu}}, \ and\ \bibinfo {author} {\bibfnamefont
  {J.}~\bibnamefont {Zhu}},\ }\href {\doibase 10.1103/PhysRevApplied.16.057001}
  {\bibfield  {journal} {\bibinfo  {journal} {Phys. Rev. Applied}\ }\textbf
  {\bibinfo {volume} {16}},\ \bibinfo {pages} {057001} (\bibinfo {year}
  {2021})}\BibitemShut {NoStop}%
\bibitem [{\citenamefont {Rosa}\ and\ \citenamefont {Ruzzene}(2020)}]{16}%
  \BibitemOpen
  \bibfield  {author} {\bibinfo {author} {\bibfnamefont {M.~I.~N.}\
  \bibnamefont {Rosa}}\ and\ \bibinfo {author} {\bibfnamefont {M.}~\bibnamefont
  {Ruzzene}},\ }\href {\doibase 10.1088/1367-2630/ab81b6} {\bibfield  {journal}
  {\bibinfo  {journal} {New Journal of Physics}\ }\textbf {\bibinfo {volume}
  {22}},\ \bibinfo {pages} {053004} (\bibinfo {year} {2020})}\BibitemShut
  {NoStop}%
\bibitem [{\citenamefont {Jiang}\ \emph {et~al.}(2019)\citenamefont {Jiang},
  \citenamefont {Mei}, \citenamefont {Zuo}, \citenamefont {Zhai}, \citenamefont
  {Li}, \citenamefont {Wen},\ and\ \citenamefont {Du}}]{17}%
  \BibitemOpen
  \bibfield  {author} {\bibinfo {author} {\bibfnamefont {Y.}~\bibnamefont
  {Jiang}}, \bibinfo {author} {\bibfnamefont {Y.}~\bibnamefont {Mei}}, \bibinfo
  {author} {\bibfnamefont {Y.}~\bibnamefont {Zuo}}, \bibinfo {author}
  {\bibfnamefont {Y.}~\bibnamefont {Zhai}}, \bibinfo {author} {\bibfnamefont
  {J.}~\bibnamefont {Li}}, \bibinfo {author} {\bibfnamefont {J.}~\bibnamefont
  {Wen}}, \ and\ \bibinfo {author} {\bibfnamefont {S.}~\bibnamefont {Du}},\
  }\href {\doibase 10.1103/PhysRevLett.123.193604} {\bibfield  {journal}
  {\bibinfo  {journal} {Phys. Rev. Lett.}\ }\textbf {\bibinfo {volume} {123}},\
  \bibinfo {pages} {193604} (\bibinfo {year} {2019})}\BibitemShut {NoStop}%
\bibitem [{\citenamefont {Gao}\ \emph {et~al.}(2020)\citenamefont {Gao},
  \citenamefont {Xue}, \citenamefont {Wang}, \citenamefont {Gu}, \citenamefont
  {Liu}, \citenamefont {Zhu},\ and\ \citenamefont {Zhang}}]{18}%
  \BibitemOpen
  \bibfield  {author} {\bibinfo {author} {\bibfnamefont {H.}~\bibnamefont
  {Gao}}, \bibinfo {author} {\bibfnamefont {H.}~\bibnamefont {Xue}}, \bibinfo
  {author} {\bibfnamefont {Q.}~\bibnamefont {Wang}}, \bibinfo {author}
  {\bibfnamefont {Z.}~\bibnamefont {Gu}}, \bibinfo {author} {\bibfnamefont
  {T.}~\bibnamefont {Liu}}, \bibinfo {author} {\bibfnamefont {J.}~\bibnamefont
  {Zhu}}, \ and\ \bibinfo {author} {\bibfnamefont {B.}~\bibnamefont {Zhang}},\
  }\href {\doibase 10.1103/PhysRevB.101.180303} {\bibfield  {journal} {\bibinfo
   {journal} {Phys. Rev. B}\ }\textbf {\bibinfo {volume} {101}},\ \bibinfo
  {pages} {180303} (\bibinfo {year} {2020})}\BibitemShut {NoStop}%
\bibitem [{\citenamefont {Staliunas}\ \emph {et~al.}(2009)\citenamefont
  {Staliunas}, \citenamefont {Herrero},\ and\ \citenamefont {Vilaseca}}]{19}%
  \BibitemOpen
  \bibfield  {author} {\bibinfo {author} {\bibfnamefont {K.}~\bibnamefont
  {Staliunas}}, \bibinfo {author} {\bibfnamefont {R.}~\bibnamefont {Herrero}},
  \ and\ \bibinfo {author} {\bibfnamefont {R.}~\bibnamefont {Vilaseca}},\
  }\href {\doibase 10.1103/PhysRevA.80.013821} {\bibfield  {journal} {\bibinfo
  {journal} {Phys. Rev. A}\ }\textbf {\bibinfo {volume} {80}},\ \bibinfo
  {pages} {013821} (\bibinfo {year} {2009})}\BibitemShut {NoStop}%
\bibitem [{\citenamefont {Botey}\ \emph {et~al.}(2010)\citenamefont {Botey},
  \citenamefont {Herrero},\ and\ \citenamefont {Staliunas}}]{20}%
  \BibitemOpen
  \bibfield  {author} {\bibinfo {author} {\bibfnamefont {M.}~\bibnamefont
  {Botey}}, \bibinfo {author} {\bibfnamefont {R.}~\bibnamefont {Herrero}}, \
  and\ \bibinfo {author} {\bibfnamefont {K.}~\bibnamefont {Staliunas}},\ }\href
  {\doibase 10.1103/PhysRevA.82.013828} {\bibfield  {journal} {\bibinfo
  {journal} {Phys. Rev. A}\ }\textbf {\bibinfo {volume} {82}},\ \bibinfo
  {pages} {013828} (\bibinfo {year} {2010})}\BibitemShut {NoStop}%
\bibitem [{\citenamefont {Hu}\ \emph {et~al.}(2021)\citenamefont {Hu},
  \citenamefont {Zhang}, \citenamefont {Zhang}, \citenamefont {Zheng},
  \citenamefont {Xiong}, \citenamefont {Yue}, \citenamefont {Wang},
  \citenamefont {Xu}, \citenamefont {Cheng}, \citenamefont {Liu},\ and\
  \citenamefont {Christensen}}]{21+1}%
  \BibitemOpen
  \bibfield  {author} {\bibinfo {author} {\bibfnamefont {B.}~\bibnamefont
  {Hu}}, \bibinfo {author} {\bibfnamefont {Z.}~\bibnamefont {Zhang}}, \bibinfo
  {author} {\bibfnamefont {H.}~\bibnamefont {Zhang}}, \bibinfo {author}
  {\bibfnamefont {L.}~\bibnamefont {Zheng}}, \bibinfo {author} {\bibfnamefont
  {W.}~\bibnamefont {Xiong}}, \bibinfo {author} {\bibfnamefont
  {Z.}~\bibnamefont {Yue}}, \bibinfo {author} {\bibfnamefont {X.}~\bibnamefont
  {Wang}}, \bibinfo {author} {\bibfnamefont {J.}~\bibnamefont {Xu}}, \bibinfo
  {author} {\bibfnamefont {Y.}~\bibnamefont {Cheng}}, \bibinfo {author}
  {\bibfnamefont {X.}~\bibnamefont {Liu}}, \ and\ \bibinfo {author}
  {\bibfnamefont {J.}~\bibnamefont {Christensen}},\ }\href {\doibase
  10.1038/s41586-021-03833-4} {\bibfield  {journal} {\bibinfo  {journal}
  {NATURE}\ }\textbf {\bibinfo {volume} {597}},\ \bibinfo {pages} {655}
  (\bibinfo {year} {2021})}\BibitemShut {NoStop}%
\bibitem [{\citenamefont {Kominis}(2015)}]{21}%
  \BibitemOpen
  \bibfield  {author} {\bibinfo {author} {\bibfnamefont {Y.}~\bibnamefont
  {Kominis}},\ }\href {\doibase 10.1103/PhysRevA.92.063849} {\bibfield
  {journal} {\bibinfo  {journal} {Phys. Rev. A}\ }\textbf {\bibinfo {volume}
  {92}},\ \bibinfo {pages} {063849} (\bibinfo {year} {2015})}\BibitemShut
  {NoStop}%
\bibitem [{\citenamefont {Bertlmann}\ \emph {et~al.}(2006)\citenamefont
  {Bertlmann}, \citenamefont {Grimus},\ and\ \citenamefont {Hiesmayr}}]{3}%
  \BibitemOpen
  \bibfield  {author} {\bibinfo {author} {\bibfnamefont {R.~A.}\ \bibnamefont
  {Bertlmann}}, \bibinfo {author} {\bibfnamefont {W.}~\bibnamefont {Grimus}}, \
  and\ \bibinfo {author} {\bibfnamefont {B.~C.}\ \bibnamefont {Hiesmayr}},\
  }\href {\doibase 10.1103/PhysRevA.73.054101} {\bibfield  {journal} {\bibinfo
  {journal} {Phys. Rev. A}\ }\textbf {\bibinfo {volume} {73}},\ \bibinfo
  {pages} {054101} (\bibinfo {year} {2006})}\BibitemShut {NoStop}%
\bibitem [{\citenamefont {Hou}\ \emph {et~al.}(2021{\natexlab{a}})\citenamefont
  {Hou}, \citenamefont {Wu},\ and\ \citenamefont {Zhang}}]{4}%
  \BibitemOpen
  \bibfield  {author} {\bibinfo {author} {\bibfnamefont {J.}~\bibnamefont
  {Hou}}, \bibinfo {author} {\bibfnamefont {Y.-J.}\ \bibnamefont {Wu}}, \ and\
  \bibinfo {author} {\bibfnamefont {C.}~\bibnamefont {Zhang}},\ }\href
  {\doibase 10.1103/PhysRevA.103.033305} {\bibfield  {journal} {\bibinfo
  {journal} {Phys. Rev. A}\ }\textbf {\bibinfo {volume} {103}},\ \bibinfo
  {pages} {033305} (\bibinfo {year} {2021}{\natexlab{a}})}\BibitemShut
  {NoStop}%
\bibitem [{\citenamefont {Tzortzakakis}\ \emph {et~al.}(2021)\citenamefont
  {Tzortzakakis}, \citenamefont {Makris}, \citenamefont {Szameit},\ and\
  \citenamefont {Economou}}]{5}%
  \BibitemOpen
  \bibfield  {author} {\bibinfo {author} {\bibfnamefont {A.~F.}\ \bibnamefont
  {Tzortzakakis}}, \bibinfo {author} {\bibfnamefont {K.~G.}\ \bibnamefont
  {Makris}}, \bibinfo {author} {\bibfnamefont {A.}~\bibnamefont {Szameit}}, \
  and\ \bibinfo {author} {\bibfnamefont {E.~N.}\ \bibnamefont {Economou}},\
  }\href {\doibase 10.1103/PhysRevResearch.3.013208} {\bibfield  {journal}
  {\bibinfo  {journal} {Phys. Rev. Research}\ }\textbf {\bibinfo {volume}
  {3}},\ \bibinfo {pages} {013208} (\bibinfo {year} {2021})}\BibitemShut
  {NoStop}%
\bibitem [{\citenamefont {Song}\ \emph {et~al.}(2019)\citenamefont {Song},
  \citenamefont {Yao},\ and\ \citenamefont {Wang}}]{6}%
  \BibitemOpen
  \bibfield  {author} {\bibinfo {author} {\bibfnamefont {F.}~\bibnamefont
  {Song}}, \bibinfo {author} {\bibfnamefont {S.}~\bibnamefont {Yao}}, \ and\
  \bibinfo {author} {\bibfnamefont {Z.}~\bibnamefont {Wang}},\ }\href {\doibase
  10.1103/PhysRevLett.123.170401} {\bibfield  {journal} {\bibinfo  {journal}
  {Phys. Rev. Lett.}\ }\textbf {\bibinfo {volume} {123}},\ \bibinfo {pages}
  {170401} (\bibinfo {year} {2019})}\BibitemShut {NoStop}%
\bibitem [{\citenamefont {Okuma}\ and\ \citenamefont {Sato}(2021)}]{7}%
  \BibitemOpen
  \bibfield  {author} {\bibinfo {author} {\bibfnamefont {N.}~\bibnamefont
  {Okuma}}\ and\ \bibinfo {author} {\bibfnamefont {M.}~\bibnamefont {Sato}},\
  }\href {\doibase 10.1103/PhysRevB.103.085428} {\bibfield  {journal} {\bibinfo
   {journal} {Phys. Rev. B}\ }\textbf {\bibinfo {volume} {103}},\ \bibinfo
  {pages} {085428} (\bibinfo {year} {2021})}\BibitemShut {NoStop}%
\bibitem [{\citenamefont {Cao}\ and\ \citenamefont {Wiersig}(2015)}]{8}%
  \BibitemOpen
  \bibfield  {author} {\bibinfo {author} {\bibfnamefont {H.}~\bibnamefont
  {Cao}}\ and\ \bibinfo {author} {\bibfnamefont {J.}~\bibnamefont {Wiersig}},\
  }\href {\doibase 10.1103/RevModPhys.87.61} {\bibfield  {journal} {\bibinfo
  {journal} {Rev. Mod. Phys.}\ }\textbf {\bibinfo {volume} {87}},\ \bibinfo
  {pages} {61} (\bibinfo {year} {2015})}\BibitemShut {NoStop}%
\bibitem [{\citenamefont {Wahlstrand}\ \emph {et~al.}(2014)\citenamefont
  {Wahlstrand}, \citenamefont {Yakimenko},\ and\ \citenamefont {Berggren}}]{9}%
  \BibitemOpen
  \bibfield  {author} {\bibinfo {author} {\bibfnamefont {B.}~\bibnamefont
  {Wahlstrand}}, \bibinfo {author} {\bibfnamefont {I.~I.}\ \bibnamefont
  {Yakimenko}}, \ and\ \bibinfo {author} {\bibfnamefont {K.-F.}\ \bibnamefont
  {Berggren}},\ }\href {\doibase 10.1103/PhysRevE.89.062910} {\bibfield
  {journal} {\bibinfo  {journal} {Phys. Rev. E}\ }\textbf {\bibinfo {volume}
  {89}},\ \bibinfo {pages} {062910} (\bibinfo {year} {2014})}\BibitemShut
  {NoStop}%
\bibitem [{\citenamefont {Fleischer}\ and\ \citenamefont
  {Moiseyev}(2005)}]{10}%
  \BibitemOpen
  \bibfield  {author} {\bibinfo {author} {\bibfnamefont {A.}~\bibnamefont
  {Fleischer}}\ and\ \bibinfo {author} {\bibfnamefont {N.}~\bibnamefont
  {Moiseyev}},\ }\href {\doibase 10.1103/PhysRevA.72.032103} {\bibfield
  {journal} {\bibinfo  {journal} {Phys. Rev. A}\ }\textbf {\bibinfo {volume}
  {72}},\ \bibinfo {pages} {032103} (\bibinfo {year} {2005})}\BibitemShut
  {NoStop}%
\bibitem [{\citenamefont {Hou}\ \emph {et~al.}(2021{\natexlab{b}})\citenamefont
  {Hou}, \citenamefont {Wu},\ and\ \citenamefont {Zhang}}]{23}%
  \BibitemOpen
  \bibfield  {author} {\bibinfo {author} {\bibfnamefont {J.}~\bibnamefont
  {Hou}}, \bibinfo {author} {\bibfnamefont {Y.-J.}\ \bibnamefont {Wu}}, \ and\
  \bibinfo {author} {\bibfnamefont {C.}~\bibnamefont {Zhang}},\ }\href
  {\doibase 10.1103/PhysRevA.103.033305} {\bibfield  {journal} {\bibinfo
  {journal} {Phys. Rev. A}\ }\textbf {\bibinfo {volume} {103}},\ \bibinfo
  {pages} {033305} (\bibinfo {year} {2021}{\natexlab{b}})}\BibitemShut
  {NoStop}%
\bibitem [{\citenamefont {Hayata}\ and\ \citenamefont {Yamamoto}(2021)}]{24}%
  \BibitemOpen
  \bibfield  {author} {\bibinfo {author} {\bibfnamefont {T.}~\bibnamefont
  {Hayata}}\ and\ \bibinfo {author} {\bibfnamefont {A.}~\bibnamefont
  {Yamamoto}},\ }\href {\doibase 10.1103/PhysRevB.104.125102} {\bibfield
  {journal} {\bibinfo  {journal} {Phys. Rev. B}\ }\textbf {\bibinfo {volume}
  {104}},\ \bibinfo {pages} {125102} (\bibinfo {year} {2021})}\BibitemShut
  {NoStop}%
\bibitem [{\citenamefont {Han}\ \emph {et~al.}(2021)\citenamefont {Han},
  \citenamefont {Liu},\ and\ \citenamefont {Liu}}]{25}%
  \BibitemOpen
  \bibfield  {author} {\bibinfo {author} {\bibfnamefont {Y.}~\bibnamefont
  {Han}}, \bibinfo {author} {\bibfnamefont {J.~S.}\ \bibnamefont {Liu}}, \ and\
  \bibinfo {author} {\bibfnamefont {C.-S.}\ \bibnamefont {Liu}},\ }\href
  {http://iopscience.iop.org/article/10.1088/1367-2630/ac3e9f} {\bibfield
  {journal} {\bibinfo  {journal} {New Journal of Physics}\ } (\bibinfo {year}
  {2021})}\BibitemShut {NoStop}%
\bibitem [{\citenamefont {Zeng}\ \emph {et~al.}(2020)\citenamefont {Zeng},
  \citenamefont {Yang},\ and\ \citenamefont {L\"u}}]{26}%
  \BibitemOpen
  \bibfield  {author} {\bibinfo {author} {\bibfnamefont {Q.-B.}\ \bibnamefont
  {Zeng}}, \bibinfo {author} {\bibfnamefont {Y.-B.}\ \bibnamefont {Yang}}, \
  and\ \bibinfo {author} {\bibfnamefont {R.}~\bibnamefont {L\"u}},\ }\href
  {\doibase 10.1103/PhysRevB.101.125418} {\bibfield  {journal} {\bibinfo
  {journal} {Phys. Rev. B}\ }\textbf {\bibinfo {volume} {101}},\ \bibinfo
  {pages} {125418} (\bibinfo {year} {2020})}\BibitemShut {NoStop}%
\bibitem [{\citenamefont {Gong}\ \emph {et~al.}(2018)\citenamefont {Gong},
  \citenamefont {Ashida}, \citenamefont {Kawabata}, \citenamefont {Takasan},
  \citenamefont {Higashikawa},\ and\ \citenamefont {Ueda}}]{27}%
  \BibitemOpen
  \bibfield  {author} {\bibinfo {author} {\bibfnamefont {Z.}~\bibnamefont
  {Gong}}, \bibinfo {author} {\bibfnamefont {Y.}~\bibnamefont {Ashida}},
  \bibinfo {author} {\bibfnamefont {K.}~\bibnamefont {Kawabata}}, \bibinfo
  {author} {\bibfnamefont {K.}~\bibnamefont {Takasan}}, \bibinfo {author}
  {\bibfnamefont {S.}~\bibnamefont {Higashikawa}}, \ and\ \bibinfo {author}
  {\bibfnamefont {M.}~\bibnamefont {Ueda}},\ }\href {\doibase
  10.1103/PhysRevX.8.031079} {\bibfield  {journal} {\bibinfo  {journal} {Phys.
  Rev. X}\ }\textbf {\bibinfo {volume} {8}},\ \bibinfo {pages} {031079}
  (\bibinfo {year} {2018})}\BibitemShut {NoStop}%
\bibitem [{\citenamefont {Guo}\ and\ \citenamefont {Chen}(2021)}]{28}%
  \BibitemOpen
  \bibfield  {author} {\bibinfo {author} {\bibfnamefont {C.-X.}\ \bibnamefont
  {Guo}}\ and\ \bibinfo {author} {\bibfnamefont {S.}~\bibnamefont {Chen}},\
  }\href {http://iopscience.iop.org/article/10.1088/1674-1056/ac3228}
  {\bibfield  {journal} {\bibinfo  {journal} {Chinese Physics B}\ } (\bibinfo
  {year} {2021})}\BibitemShut {NoStop}%
\bibitem [{\citenamefont {Longhi}(2021)}]{29}%
  \BibitemOpen
  \bibfield  {author} {\bibinfo {author} {\bibfnamefont {S.}~\bibnamefont
  {Longhi}},\ }\href {\doibase 10.1103/PhysRevB.103.054203} {\bibfield
  {journal} {\bibinfo  {journal} {Phys. Rev. B}\ }\textbf {\bibinfo {volume}
  {103}},\ \bibinfo {pages} {054203} (\bibinfo {year} {2021})}\BibitemShut
  {NoStop}%
\bibitem [{\citenamefont {Yao}\ and\ \citenamefont {Wang}(2018)}]{30}%
  \BibitemOpen
  \bibfield  {author} {\bibinfo {author} {\bibfnamefont {S.}~\bibnamefont
  {Yao}}\ and\ \bibinfo {author} {\bibfnamefont {Z.}~\bibnamefont {Wang}},\
  }\href {\doibase 10.1103/PhysRevLett.121.086803} {\bibfield  {journal}
  {\bibinfo  {journal} {Phys. Rev. Lett.}\ }\textbf {\bibinfo {volume} {121}},\
  \bibinfo {pages} {086803} (\bibinfo {year} {2018})}\BibitemShut {NoStop}%
\bibitem [{\citenamefont {Yao}\ \emph {et~al.}(2018)\citenamefont {Yao},
  \citenamefont {Song},\ and\ \citenamefont {Wang}}]{31}%
  \BibitemOpen
  \bibfield  {author} {\bibinfo {author} {\bibfnamefont {S.}~\bibnamefont
  {Yao}}, \bibinfo {author} {\bibfnamefont {F.}~\bibnamefont {Song}}, \ and\
  \bibinfo {author} {\bibfnamefont {Z.}~\bibnamefont {Wang}},\ }\href {\doibase
  10.1103/PhysRevLett.121.136802} {\bibfield  {journal} {\bibinfo  {journal}
  {Phys. Rev. Lett.}\ }\textbf {\bibinfo {volume} {121}},\ \bibinfo {pages}
  {136802} (\bibinfo {year} {2018})}\BibitemShut {NoStop}%
\bibitem [{\citenamefont {Lee}\ \emph {et~al.}(2020)\citenamefont {Lee},
  \citenamefont {Lee},\ and\ \citenamefont {Yang}}]{32}%
  \BibitemOpen
  \bibfield  {author} {\bibinfo {author} {\bibfnamefont {E.}~\bibnamefont
  {Lee}}, \bibinfo {author} {\bibfnamefont {H.}~\bibnamefont {Lee}}, \ and\
  \bibinfo {author} {\bibfnamefont {B.-J.}\ \bibnamefont {Yang}},\ }\href
  {\doibase 10.1103/PhysRevB.101.121109} {\bibfield  {journal} {\bibinfo
  {journal} {Phys. Rev. B}\ }\textbf {\bibinfo {volume} {101}},\ \bibinfo
  {pages} {121109} (\bibinfo {year} {2020})}\BibitemShut {NoStop}%
\bibitem [{\citenamefont {Li}\ \emph {et~al.}(2020)\citenamefont {Li},
  \citenamefont {Liu}, \citenamefont {Li},\ and\ \citenamefont {Liu}}]{33}%
  \BibitemOpen
  \bibfield  {author} {\bibinfo {author} {\bibfnamefont {S.}~\bibnamefont
  {Li}}, \bibinfo {author} {\bibfnamefont {M.}~\bibnamefont {Liu}}, \bibinfo
  {author} {\bibfnamefont {F.}~\bibnamefont {Li}}, \ and\ \bibinfo {author}
  {\bibfnamefont {B.}~\bibnamefont {Liu}},\ }\href {\doibase
  10.1088/1402-4896/abc580} {\bibfield  {journal} {\bibinfo  {journal} {Physica
  Scripta}\ }\textbf {\bibinfo {volume} {96}},\ \bibinfo {pages} {015402}
  (\bibinfo {year} {2020})}\BibitemShut {NoStop}%
\bibitem [{\citenamefont {Entin-Wohlman}\ and\ \citenamefont
  {Aharony}(2019)}]{34}%
  \BibitemOpen
  \bibfield  {author} {\bibinfo {author} {\bibfnamefont {O.}~\bibnamefont
  {Entin-Wohlman}}\ and\ \bibinfo {author} {\bibfnamefont {A.}~\bibnamefont
  {Aharony}},\ }\href {\doibase 10.1103/PhysRevResearch.1.033112} {\bibfield
  {journal} {\bibinfo  {journal} {Phys. Rev. Research}\ }\textbf {\bibinfo
  {volume} {1}},\ \bibinfo {pages} {033112} (\bibinfo {year}
  {2019})}\BibitemShut {NoStop}%
\bibitem [{\citenamefont {Aligia}\ \emph {et~al.}(2000)\citenamefont {Aligia},
  \citenamefont {Hallberg}, \citenamefont {Batista},\ and\ \citenamefont
  {Ortiz}}]{35}%
  \BibitemOpen
  \bibfield  {author} {\bibinfo {author} {\bibfnamefont {A.~A.}\ \bibnamefont
  {Aligia}}, \bibinfo {author} {\bibfnamefont {K.}~\bibnamefont {Hallberg}},
  \bibinfo {author} {\bibfnamefont {C.~D.}\ \bibnamefont {Batista}}, \ and\
  \bibinfo {author} {\bibfnamefont {G.}~\bibnamefont {Ortiz}},\ }\href
  {\doibase 10.1103/PhysRevB.61.7883} {\bibfield  {journal} {\bibinfo
  {journal} {Phys. Rev. B}\ }\textbf {\bibinfo {volume} {61}},\ \bibinfo
  {pages} {7883} (\bibinfo {year} {2000})}\BibitemShut {NoStop}%
\bibitem [{\citenamefont {Sil}\ and\ \citenamefont {Ghosh}(2019)}]{36}%
  \BibitemOpen
  \bibfield  {author} {\bibinfo {author} {\bibfnamefont {A.}~\bibnamefont
  {Sil}}\ and\ \bibinfo {author} {\bibfnamefont {A.~K.}\ \bibnamefont
  {Ghosh}},\ }\href {\doibase 10.1088/1361-648x/ab4750} {\bibfield  {journal}
  {\bibinfo  {journal} {Journal of Physics: Condensed Matter}\ }\textbf
  {\bibinfo {volume} {32}},\ \bibinfo {pages} {025601} (\bibinfo {year}
  {2019})}\BibitemShut {NoStop}%
\bibitem [{\citenamefont {Zuo}\ \emph {et~al.}(2021)\citenamefont {Zuo},
  \citenamefont {Benalcazar}, \citenamefont {Liu},\ and\ \citenamefont
  {Liu}}]{37}%
  \BibitemOpen
  \bibfield  {author} {\bibinfo {author} {\bibfnamefont {Z.-W.}\ \bibnamefont
  {Zuo}}, \bibinfo {author} {\bibfnamefont {W.~A.}\ \bibnamefont {Benalcazar}},
  \bibinfo {author} {\bibfnamefont {Y.}~\bibnamefont {Liu}}, \ and\ \bibinfo
  {author} {\bibfnamefont {C.-X.}\ \bibnamefont {Liu}},\ }\href {\doibase
  10.1088/1361-6463/ac12f7} {\bibfield  {journal} {\bibinfo  {journal} {Journal
  of Physics D: Applied Physics}\ }\textbf {\bibinfo {volume} {54}},\ \bibinfo
  {pages} {414004} (\bibinfo {year} {2021})}\BibitemShut {NoStop}%
\bibitem [{\citenamefont {Kuno}\ \emph {et~al.}(2015)\citenamefont {Kuno},
  \citenamefont {Nakafuji},\ and\ \citenamefont {Ichinose}}]{38}%
  \BibitemOpen
  \bibfield  {author} {\bibinfo {author} {\bibfnamefont {Y.}~\bibnamefont
  {Kuno}}, \bibinfo {author} {\bibfnamefont {T.}~\bibnamefont {Nakafuji}}, \
  and\ \bibinfo {author} {\bibfnamefont {I.}~\bibnamefont {Ichinose}},\ }\href
  {\doibase 10.1103/PhysRevA.92.063630} {\bibfield  {journal} {\bibinfo
  {journal} {Phys. Rev. A}\ }\textbf {\bibinfo {volume} {92}},\ \bibinfo
  {pages} {063630} (\bibinfo {year} {2015})}\BibitemShut {NoStop}%
\bibitem [{\citenamefont {Beugeling}\ \emph {et~al.}(2012)\citenamefont
  {Beugeling}, \citenamefont {Everts},\ and\ \citenamefont
  {Morais~Smith}}]{39}%
  \BibitemOpen
  \bibfield  {author} {\bibinfo {author} {\bibfnamefont {W.}~\bibnamefont
  {Beugeling}}, \bibinfo {author} {\bibfnamefont {J.~C.}\ \bibnamefont
  {Everts}}, \ and\ \bibinfo {author} {\bibfnamefont {C.}~\bibnamefont
  {Morais~Smith}},\ }\href {\doibase 10.1103/PhysRevB.86.195129} {\bibfield
  {journal} {\bibinfo  {journal} {Phys. Rev. B}\ }\textbf {\bibinfo {volume}
  {86}},\ \bibinfo {pages} {195129} (\bibinfo {year} {2012})}\BibitemShut
  {NoStop}%
\bibitem [{\citenamefont {Goldman}\ \emph {et~al.}(2013)\citenamefont
  {Goldman}, \citenamefont {Anisimovas}, \citenamefont {Gerbier}, \citenamefont
  {Öhberg}, \citenamefont {Spielman},\ and\ \citenamefont
  {Juzeli{\={u}}nas}}]{40}%
  \BibitemOpen
  \bibfield  {author} {\bibinfo {author} {\bibfnamefont {N.}~\bibnamefont
  {Goldman}}, \bibinfo {author} {\bibfnamefont {E.}~\bibnamefont {Anisimovas}},
  \bibinfo {author} {\bibfnamefont {F.}~\bibnamefont {Gerbier}}, \bibinfo
  {author} {\bibfnamefont {P.}~\bibnamefont {Öhberg}}, \bibinfo {author}
  {\bibfnamefont {I.~B.}\ \bibnamefont {Spielman}}, \ and\ \bibinfo {author}
  {\bibfnamefont {G.}~\bibnamefont {Juzeli{\={u}}nas}},\ }\href {\doibase
  10.1088/1367-2630/15/1/013025} {\bibfield  {journal} {\bibinfo  {journal}
  {New Journal of Physics}\ }\textbf {\bibinfo {volume} {15}},\ \bibinfo
  {pages} {013025} (\bibinfo {year} {2013})}\BibitemShut {NoStop}%
\bibitem [{\citenamefont {Haldane}(1988)}]{41}%
  \BibitemOpen
  \bibfield  {author} {\bibinfo {author} {\bibfnamefont {F.~D.~M.}\
  \bibnamefont {Haldane}},\ }\href {\doibase 10.1103/PhysRevLett.61.2015}
  {\bibfield  {journal} {\bibinfo  {journal} {Phys. Rev. Lett.}\ }\textbf
  {\bibinfo {volume} {61}},\ \bibinfo {pages} {2015} (\bibinfo {year}
  {1988})}\BibitemShut {NoStop}%
\bibitem [{\citenamefont {Bhat}\ \emph {et~al.}(2007)\citenamefont {Bhat},
  \citenamefont {Kr\"amer}, \citenamefont {Cooper},\ and\ \citenamefont
  {Holland}}]{43}%
  \BibitemOpen
  \bibfield  {author} {\bibinfo {author} {\bibfnamefont {R.}~\bibnamefont
  {Bhat}}, \bibinfo {author} {\bibfnamefont {M.}~\bibnamefont {Kr\"amer}},
  \bibinfo {author} {\bibfnamefont {J.}~\bibnamefont {Cooper}}, \ and\ \bibinfo
  {author} {\bibfnamefont {M.~J.}\ \bibnamefont {Holland}},\ }\href {\doibase
  10.1103/PhysRevA.76.043601} {\bibfield  {journal} {\bibinfo  {journal} {Phys.
  Rev. A}\ }\textbf {\bibinfo {volume} {76}},\ \bibinfo {pages} {043601}
  (\bibinfo {year} {2007})}\BibitemShut {NoStop}%
\bibitem [{\citenamefont {Zheng}\ \emph {et~al.}(2011)\citenamefont {Zheng},
  \citenamefont {Zhang},\ and\ \citenamefont {Wu}}]{44}%
  \BibitemOpen
  \bibfield  {author} {\bibinfo {author} {\bibfnamefont {D.}~\bibnamefont
  {Zheng}}, \bibinfo {author} {\bibfnamefont {G.-M.}\ \bibnamefont {Zhang}}, \
  and\ \bibinfo {author} {\bibfnamefont {C.}~\bibnamefont {Wu}},\ }\href
  {\doibase 10.1103/PhysRevB.84.205121} {\bibfield  {journal} {\bibinfo
  {journal} {Phys. Rev. B}\ }\textbf {\bibinfo {volume} {84}},\ \bibinfo
  {pages} {205121} (\bibinfo {year} {2011})}\BibitemShut {NoStop}%
\bibitem [{\citenamefont {Choi}\ and\ \citenamefont {Doniach}(1985)}]{46}%
  \BibitemOpen
  \bibfield  {author} {\bibinfo {author} {\bibfnamefont {M.~Y.}\ \bibnamefont
  {Choi}}\ and\ \bibinfo {author} {\bibfnamefont {S.}~\bibnamefont {Doniach}},\
  }\href {\doibase 10.1103/PhysRevB.31.4516} {\bibfield  {journal} {\bibinfo
  {journal} {Phys. Rev. B}\ }\textbf {\bibinfo {volume} {31}},\ \bibinfo
  {pages} {4516} (\bibinfo {year} {1985})}\BibitemShut {NoStop}%
\bibitem [{\citenamefont {Xu}\ \emph {et~al.}(2021)\citenamefont {Xu},
  \citenamefont {Zhang}, \citenamefont {Luo}, \citenamefont {Yu}, \citenamefont
  {Li},\ and\ \citenamefont {Zhang}}]{64}%
  \BibitemOpen
  \bibfield  {author} {\bibinfo {author} {\bibfnamefont {K.}~\bibnamefont
  {Xu}}, \bibinfo {author} {\bibfnamefont {X.}~\bibnamefont {Zhang}}, \bibinfo
  {author} {\bibfnamefont {K.}~\bibnamefont {Luo}}, \bibinfo {author}
  {\bibfnamefont {R.}~\bibnamefont {Yu}}, \bibinfo {author} {\bibfnamefont
  {D.}~\bibnamefont {Li}}, \ and\ \bibinfo {author} {\bibfnamefont
  {H.}~\bibnamefont {Zhang}},\ }\href {\doibase 10.1103/PhysRevB.103.125411}
  {\bibfield  {journal} {\bibinfo  {journal} {Phys. Rev. B}\ }\textbf {\bibinfo
  {volume} {103}},\ \bibinfo {pages} {125411} (\bibinfo {year}
  {2021})}\BibitemShut {NoStop}%
\bibitem [{\citenamefont {Qi}\ \emph {et~al.}(2021)\citenamefont {Qi},
  \citenamefont {Yan}, \citenamefont {Xing}, \citenamefont {Zhao},
  \citenamefont {Liu}, \citenamefont {Cui}, \citenamefont {Han}, \citenamefont
  {Zhang},\ and\ \citenamefont {Wang}}]{65}%
  \BibitemOpen
  \bibfield  {author} {\bibinfo {author} {\bibfnamefont {L.}~\bibnamefont
  {Qi}}, \bibinfo {author} {\bibfnamefont {Y.}~\bibnamefont {Yan}}, \bibinfo
  {author} {\bibfnamefont {Y.}~\bibnamefont {Xing}}, \bibinfo {author}
  {\bibfnamefont {X.-D.}\ \bibnamefont {Zhao}}, \bibinfo {author}
  {\bibfnamefont {S.}~\bibnamefont {Liu}}, \bibinfo {author} {\bibfnamefont
  {W.-X.}\ \bibnamefont {Cui}}, \bibinfo {author} {\bibfnamefont
  {X.}~\bibnamefont {Han}}, \bibinfo {author} {\bibfnamefont {S.}~\bibnamefont
  {Zhang}}, \ and\ \bibinfo {author} {\bibfnamefont {H.-F.}\ \bibnamefont
  {Wang}},\ }\href {\doibase 10.1103/PhysRevResearch.3.023037} {\bibfield
  {journal} {\bibinfo  {journal} {Phys. Rev. Research}\ }\textbf {\bibinfo
  {volume} {3}},\ \bibinfo {pages} {023037} (\bibinfo {year}
  {2021})}\BibitemShut {NoStop}%
\bibitem [{\citenamefont {Lin}\ \emph {et~al.}(2016)\citenamefont {Lin},
  \citenamefont {Zhang}, \citenamefont {Li},\ and\ \citenamefont {Song}}]{66}%
  \BibitemOpen
  \bibfield  {author} {\bibinfo {author} {\bibfnamefont {S.}~\bibnamefont
  {Lin}}, \bibinfo {author} {\bibfnamefont {X.~Z.}\ \bibnamefont {Zhang}},
  \bibinfo {author} {\bibfnamefont {C.}~\bibnamefont {Li}}, \ and\ \bibinfo
  {author} {\bibfnamefont {Z.}~\bibnamefont {Song}},\ }\href {\doibase
  10.1103/PhysRevA.94.042133} {\bibfield  {journal} {\bibinfo  {journal} {Phys.
  Rev. A}\ }\textbf {\bibinfo {volume} {94}},\ \bibinfo {pages} {042133}
  (\bibinfo {year} {2016})}\BibitemShut {NoStop}%
\bibitem [{\citenamefont {Ch\'avez}\ \emph {et~al.}(2021)\citenamefont
  {Ch\'avez}, \citenamefont {Mattiotti}, \citenamefont {M\'endez-Berm\'udez},
  \citenamefont {Borgonovi},\ and\ \citenamefont {Celardo}}]{67}%
  \BibitemOpen
  \bibfield  {author} {\bibinfo {author} {\bibfnamefont {N.~C.}\ \bibnamefont
  {Ch\'avez}}, \bibinfo {author} {\bibfnamefont {F.}~\bibnamefont {Mattiotti}},
  \bibinfo {author} {\bibfnamefont {J.~A.}\ \bibnamefont
  {M\'endez-Berm\'udez}}, \bibinfo {author} {\bibfnamefont {F.}~\bibnamefont
  {Borgonovi}}, \ and\ \bibinfo {author} {\bibfnamefont {G.~L.}\ \bibnamefont
  {Celardo}},\ }\href {\doibase 10.1103/PhysRevLett.126.153201} {\bibfield
  {journal} {\bibinfo  {journal} {Phys. Rev. Lett.}\ }\textbf {\bibinfo
  {volume} {126}},\ \bibinfo {pages} {153201} (\bibinfo {year}
  {2021})}\BibitemShut {NoStop}%
\bibitem [{\citenamefont {Zhang}\ and\ \citenamefont {Song}(2020)}]{68}%
  \BibitemOpen
  \bibfield  {author} {\bibinfo {author} {\bibfnamefont {X.~Z.}\ \bibnamefont
  {Zhang}}\ and\ \bibinfo {author} {\bibfnamefont {Z.}~\bibnamefont {Song}},\
  }\href {\doibase 10.1103/PhysRevB.102.174303} {\bibfield  {journal} {\bibinfo
   {journal} {Phys. Rev. B}\ }\textbf {\bibinfo {volume} {102}},\ \bibinfo
  {pages} {174303} (\bibinfo {year} {2020})}\BibitemShut {NoStop}%
\bibitem [{\citenamefont {P\'erez-Gonz\'alez}\ \emph
  {et~al.}(2019)\citenamefont {P\'erez-Gonz\'alez}, \citenamefont {Bello},
  \citenamefont {G\'omez-Le\'on},\ and\ \citenamefont {Platero}}]{69}%
  \BibitemOpen
  \bibfield  {author} {\bibinfo {author} {\bibfnamefont {B.}~\bibnamefont
  {P\'erez-Gonz\'alez}}, \bibinfo {author} {\bibfnamefont {M.}~\bibnamefont
  {Bello}}, \bibinfo {author} {\bibfnamefont {A.}~\bibnamefont
  {G\'omez-Le\'on}}, \ and\ \bibinfo {author} {\bibfnamefont {G.}~\bibnamefont
  {Platero}},\ }\href {\doibase 10.1103/PhysRevB.99.035146} {\bibfield
  {journal} {\bibinfo  {journal} {Phys. Rev. B}\ }\textbf {\bibinfo {volume}
  {99}},\ \bibinfo {pages} {035146} (\bibinfo {year} {2019})}\BibitemShut
  {NoStop}%
\bibitem [{\citenamefont {Lee}(2016)}]{70}%
  \BibitemOpen
  \bibfield  {author} {\bibinfo {author} {\bibfnamefont {T.~E.}\ \bibnamefont
  {Lee}},\ }\href {\doibase 10.1103/PhysRevLett.116.133903} {\bibfield
  {journal} {\bibinfo  {journal} {Phys. Rev. Lett.}\ }\textbf {\bibinfo
  {volume} {116}},\ \bibinfo {pages} {133903} (\bibinfo {year}
  {2016})}\BibitemShut {NoStop}%
\bibitem [{\citenamefont {Bao}\ \emph {et~al.}(2021)\citenamefont {Bao},
  \citenamefont {Guo},\ and\ \citenamefont {Tan}}]{70-1}%
  \BibitemOpen
  \bibfield  {author} {\bibinfo {author} {\bibfnamefont {X.-X.}\ \bibnamefont
  {Bao}}, \bibinfo {author} {\bibfnamefont {G.-F.}\ \bibnamefont {Guo}}, \ and\
  \bibinfo {author} {\bibfnamefont {L.}~\bibnamefont {Tan}},\ }\href {\doibase
  10.1088/1361-648x/ac2040} {\bibfield  {journal} {\bibinfo  {journal} {Journal
  of Physics: Condensed Matter}\ }\textbf {\bibinfo {volume} {33}},\ \bibinfo
  {pages} {465403} (\bibinfo {year} {2021})}\BibitemShut {NoStop}%
\bibitem [{\citenamefont {Hsu}\ and\ \citenamefont {Chen}(2020)}]{701}%
  \BibitemOpen
  \bibfield  {author} {\bibinfo {author} {\bibfnamefont {H.-C.}\ \bibnamefont
  {Hsu}}\ and\ \bibinfo {author} {\bibfnamefont {T.-W.}\ \bibnamefont {Chen}},\
  }\href {\doibase 10.1103/PhysRevB.102.205425} {\bibfield  {journal} {\bibinfo
   {journal} {Phys. Rev. B}\ }\textbf {\bibinfo {volume} {102}},\ \bibinfo
  {pages} {205425} (\bibinfo {year} {2020})}\BibitemShut {NoStop}%
\bibitem [{\citenamefont {Grossmann}\ \emph {et~al.}(1991)\citenamefont
  {Grossmann}, \citenamefont {Dittrich}, \citenamefont {Jung},\ and\
  \citenamefont {H\"anggi}}]{47}%
  \BibitemOpen
  \bibfield  {author} {\bibinfo {author} {\bibfnamefont {F.}~\bibnamefont
  {Grossmann}}, \bibinfo {author} {\bibfnamefont {T.}~\bibnamefont {Dittrich}},
  \bibinfo {author} {\bibfnamefont {P.}~\bibnamefont {Jung}}, \ and\ \bibinfo
  {author} {\bibfnamefont {P.}~\bibnamefont {H\"anggi}},\ }\href {\doibase
  10.1103/PhysRevLett.67.516} {\bibfield  {journal} {\bibinfo  {journal} {Phys.
  Rev. Lett.}\ }\textbf {\bibinfo {volume} {67}},\ \bibinfo {pages} {516}
  (\bibinfo {year} {1991})}\BibitemShut {NoStop}%
\bibitem [{\citenamefont {Luo}\ \emph {et~al.}(2014)\citenamefont {Luo},
  \citenamefont {Li}, \citenamefont {You},\ and\ \citenamefont {Wu}}]{48}%
  \BibitemOpen
  \bibfield  {author} {\bibinfo {author} {\bibfnamefont {X.}~\bibnamefont
  {Luo}}, \bibinfo {author} {\bibfnamefont {L.}~\bibnamefont {Li}}, \bibinfo
  {author} {\bibfnamefont {L.}~\bibnamefont {You}}, \ and\ \bibinfo {author}
  {\bibfnamefont {B.}~\bibnamefont {Wu}},\ }\href {\doibase
  10.1088/1367-2630/16/1/013007} {\bibfield  {journal} {\bibinfo  {journal}
  {New Journal of Physics}\ }\textbf {\bibinfo {volume} {16}},\ \bibinfo
  {pages} {013007} (\bibinfo {year} {2014})}\BibitemShut {NoStop}%
\bibitem [{\citenamefont {Lignier}\ \emph {et~al.}(2007)\citenamefont
  {Lignier}, \citenamefont {Sias}, \citenamefont {Ciampini}, \citenamefont
  {Singh}, \citenamefont {Zenesini}, \citenamefont {Morsch},\ and\
  \citenamefont {Arimondo}}]{49}%
  \BibitemOpen
  \bibfield  {author} {\bibinfo {author} {\bibfnamefont {H.}~\bibnamefont
  {Lignier}}, \bibinfo {author} {\bibfnamefont {C.}~\bibnamefont {Sias}},
  \bibinfo {author} {\bibfnamefont {D.}~\bibnamefont {Ciampini}}, \bibinfo
  {author} {\bibfnamefont {Y.}~\bibnamefont {Singh}}, \bibinfo {author}
  {\bibfnamefont {A.}~\bibnamefont {Zenesini}}, \bibinfo {author}
  {\bibfnamefont {O.}~\bibnamefont {Morsch}}, \ and\ \bibinfo {author}
  {\bibfnamefont {E.}~\bibnamefont {Arimondo}},\ }\href {\doibase
  10.1103/PhysRevLett.99.220403} {\bibfield  {journal} {\bibinfo  {journal}
  {Phys. Rev. Lett.}\ }\textbf {\bibinfo {volume} {99}},\ \bibinfo {pages}
  {220403} (\bibinfo {year} {2007})}\BibitemShut {NoStop}%
\bibitem [{\citenamefont {Zhou}\ \emph {et~al.}(2009)\citenamefont {Zhou},
  \citenamefont {Yang}, \citenamefont {Liu}, \citenamefont {Sun},\ and\
  \citenamefont {Nori}}]{50}%
  \BibitemOpen
  \bibfield  {author} {\bibinfo {author} {\bibfnamefont {L.}~\bibnamefont
  {Zhou}}, \bibinfo {author} {\bibfnamefont {S.}~\bibnamefont {Yang}}, \bibinfo
  {author} {\bibfnamefont {Y.-x.}\ \bibnamefont {Liu}}, \bibinfo {author}
  {\bibfnamefont {C.~P.}\ \bibnamefont {Sun}}, \ and\ \bibinfo {author}
  {\bibfnamefont {F.}~\bibnamefont {Nori}},\ }\href {\doibase
  10.1103/PhysRevA.80.062109} {\bibfield  {journal} {\bibinfo  {journal} {Phys.
  Rev. A}\ }\textbf {\bibinfo {volume} {80}},\ \bibinfo {pages} {062109}
  (\bibinfo {year} {2009})}\BibitemShut {NoStop}%
\bibitem [{\citenamefont {Zhou}\ and\ \citenamefont {Kuang}(2010)}]{51}%
  \BibitemOpen
  \bibfield  {author} {\bibinfo {author} {\bibfnamefont {L.}~\bibnamefont
  {Zhou}}\ and\ \bibinfo {author} {\bibfnamefont {L.-M.}\ \bibnamefont
  {Kuang}},\ }\href {\doibase 10.1103/PhysRevA.82.042113} {\bibfield  {journal}
  {\bibinfo  {journal} {Phys. Rev. A}\ }\textbf {\bibinfo {volume} {82}},\
  \bibinfo {pages} {042113} (\bibinfo {year} {2010})}\BibitemShut {NoStop}%
\bibitem [{\citenamefont {Hauke}\ \emph {et~al.}(2012)\citenamefont {Hauke},
  \citenamefont {Tieleman}, \citenamefont {Celi}, \citenamefont
  {\"Olschl\"ager}, \citenamefont {Simonet}, \citenamefont {Struck},
  \citenamefont {Weinberg}, \citenamefont {Windpassinger}, \citenamefont
  {Sengstock}, \citenamefont {Lewenstein},\ and\ \citenamefont {Eckardt}}]{53}%
  \BibitemOpen
  \bibfield  {author} {\bibinfo {author} {\bibfnamefont {P.}~\bibnamefont
  {Hauke}}, \bibinfo {author} {\bibfnamefont {O.}~\bibnamefont {Tieleman}},
  \bibinfo {author} {\bibfnamefont {A.}~\bibnamefont {Celi}}, \bibinfo {author}
  {\bibfnamefont {C.}~\bibnamefont {\"Olschl\"ager}}, \bibinfo {author}
  {\bibfnamefont {J.}~\bibnamefont {Simonet}}, \bibinfo {author} {\bibfnamefont
  {J.}~\bibnamefont {Struck}}, \bibinfo {author} {\bibfnamefont
  {M.}~\bibnamefont {Weinberg}}, \bibinfo {author} {\bibfnamefont
  {P.}~\bibnamefont {Windpassinger}}, \bibinfo {author} {\bibfnamefont
  {K.}~\bibnamefont {Sengstock}}, \bibinfo {author} {\bibfnamefont
  {M.}~\bibnamefont {Lewenstein}}, \ and\ \bibinfo {author} {\bibfnamefont
  {A.}~\bibnamefont {Eckardt}},\ }\href {\doibase
  10.1103/PhysRevLett.109.145301} {\bibfield  {journal} {\bibinfo  {journal}
  {Phys. Rev. Lett.}\ }\textbf {\bibinfo {volume} {109}},\ \bibinfo {pages}
  {145301} (\bibinfo {year} {2012})}\BibitemShut {NoStop}%
\bibitem [{\citenamefont {Meinert}\ \emph {et~al.}(2016)\citenamefont
  {Meinert}, \citenamefont {Mark}, \citenamefont {Lauber}, \citenamefont
  {Daley},\ and\ \citenamefont {N\"agerl}}]{54}%
  \BibitemOpen
  \bibfield  {author} {\bibinfo {author} {\bibfnamefont {F.}~\bibnamefont
  {Meinert}}, \bibinfo {author} {\bibfnamefont {M.~J.}\ \bibnamefont {Mark}},
  \bibinfo {author} {\bibfnamefont {K.}~\bibnamefont {Lauber}}, \bibinfo
  {author} {\bibfnamefont {A.~J.}\ \bibnamefont {Daley}}, \ and\ \bibinfo
  {author} {\bibfnamefont {H.-C.}\ \bibnamefont {N\"agerl}},\ }\href {\doibase
  10.1103/PhysRevLett.116.205301} {\bibfield  {journal} {\bibinfo  {journal}
  {Phys. Rev. Lett.}\ }\textbf {\bibinfo {volume} {116}},\ \bibinfo {pages}
  {205301} (\bibinfo {year} {2016})}\BibitemShut {NoStop}%
\bibitem [{\citenamefont {Eckardt}(2017)}]{55}%
  \BibitemOpen
  \bibfield  {author} {\bibinfo {author} {\bibfnamefont {A.}~\bibnamefont
  {Eckardt}},\ }\href {\doibase 10.1103/RevModPhys.89.011004} {\bibfield
  {journal} {\bibinfo  {journal} {Rev. Mod. Phys.}\ }\textbf {\bibinfo {volume}
  {89}},\ \bibinfo {pages} {011004} (\bibinfo {year} {2017})}\BibitemShut
  {NoStop}%
\bibitem [{\citenamefont {Holthaus}(2015)}]{atom1}%
  \BibitemOpen
  \bibfield  {author} {\bibinfo {author} {\bibfnamefont {M.}~\bibnamefont
  {Holthaus}},\ }\href {\doibase 10.1088/0953-4075/49/1/013001} {\bibfield
  {journal} {\bibinfo  {journal} {Journal of Physics B: Atomic, Molecular and
  Optical Physics}\ }\textbf {\bibinfo {volume} {49}},\ \bibinfo {pages}
  {013001} (\bibinfo {year} {2015})}\BibitemShut {NoStop}%
\bibitem [{\citenamefont {Dauphin}\ \emph {et~al.}(2017)\citenamefont
  {Dauphin}, \citenamefont {Tran}, \citenamefont {Lewenstein},\ and\
  \citenamefont {Goldman}}]{atom2}%
  \BibitemOpen
  \bibfield  {author} {\bibinfo {author} {\bibfnamefont {A.}~\bibnamefont
  {Dauphin}}, \bibinfo {author} {\bibfnamefont {D.-T.}\ \bibnamefont {Tran}},
  \bibinfo {author} {\bibfnamefont {M.}~\bibnamefont {Lewenstein}}, \ and\
  \bibinfo {author} {\bibfnamefont {N.}~\bibnamefont {Goldman}},\ }\href
  {\doibase 10.1088/2053-1583/aa6a3b} {\bibfield  {journal} {\bibinfo
  {journal} {2D Materials}\ }\textbf {\bibinfo {volume} {4}},\ \bibinfo {pages}
  {024010} (\bibinfo {year} {2017})}\BibitemShut {NoStop}%
\bibitem [{\citenamefont {Quelle}\ \emph {et~al.}(2017)\citenamefont {Quelle},
  \citenamefont {Weitenberg}, \citenamefont {Sengstock},\ and\ \citenamefont
  {Smith}}]{atom3}%
  \BibitemOpen
  \bibfield  {author} {\bibinfo {author} {\bibfnamefont {A.}~\bibnamefont
  {Quelle}}, \bibinfo {author} {\bibfnamefont {C.}~\bibnamefont {Weitenberg}},
  \bibinfo {author} {\bibfnamefont {K.}~\bibnamefont {Sengstock}}, \ and\
  \bibinfo {author} {\bibfnamefont {C.~M.}\ \bibnamefont {Smith}},\ }\href
  {\doibase 10.1088/1367-2630/aa8646} {\bibfield  {journal} {\bibinfo
  {journal} {New Journal of Physics}\ }\textbf {\bibinfo {volume} {19}},\
  \bibinfo {pages} {113010} (\bibinfo {year} {2017})}\BibitemShut {NoStop}%
\bibitem [{\citenamefont {Rechtsman}\ \emph {et~al.}(2013)\citenamefont
  {Rechtsman}, \citenamefont {Zeuner}, \citenamefont {Plotnik}, \citenamefont
  {Lumer}, \citenamefont {Podolsky}, \citenamefont {Dreisow}, \citenamefont
  {Nolte}, \citenamefont {Segev},\ and\ \citenamefont {Szameit}}]{56}%
  \BibitemOpen
  \bibfield  {author} {\bibinfo {author} {\bibfnamefont {M.}~\bibnamefont
  {Rechtsman}}, \bibinfo {author} {\bibfnamefont {J.}~\bibnamefont {Zeuner}},
  \bibinfo {author} {\bibfnamefont {Y.}~\bibnamefont {Plotnik}}, \bibinfo
  {author} {\bibfnamefont {Y.}~\bibnamefont {Lumer}}, \bibinfo {author}
  {\bibfnamefont {D.}~\bibnamefont {Podolsky}}, \bibinfo {author}
  {\bibfnamefont {F.}~\bibnamefont {Dreisow}}, \bibinfo {author} {\bibfnamefont
  {S.}~\bibnamefont {Nolte}}, \bibinfo {author} {\bibfnamefont
  {M.}~\bibnamefont {Segev}}, \ and\ \bibinfo {author} {\bibfnamefont
  {A.}~\bibnamefont {Szameit}},\ }\href {\doibase 10.1038/nature12066}
  {\bibfield  {journal} {\bibinfo  {journal} {NATURE}\ }\textbf {\bibinfo
  {volume} {496}},\ \bibinfo {pages} {196} (\bibinfo {year}
  {2013})}\BibitemShut {NoStop}%
\bibitem [{\citenamefont {Cheng}\ \emph {et~al.}(2019)\citenamefont {Cheng},
  \citenamefont {Pan}, \citenamefont {Wang}, \citenamefont {Zhang},
  \citenamefont {Yu}, \citenamefont {Gover}, \citenamefont {Zhang},
  \citenamefont {Li}, \citenamefont {Zhou},\ and\ \citenamefont {Zhu}}]{57}%
  \BibitemOpen
  \bibfield  {author} {\bibinfo {author} {\bibfnamefont {Q.}~\bibnamefont
  {Cheng}}, \bibinfo {author} {\bibfnamefont {Y.}~\bibnamefont {Pan}}, \bibinfo
  {author} {\bibfnamefont {H.}~\bibnamefont {Wang}}, \bibinfo {author}
  {\bibfnamefont {C.}~\bibnamefont {Zhang}}, \bibinfo {author} {\bibfnamefont
  {D.}~\bibnamefont {Yu}}, \bibinfo {author} {\bibfnamefont {A.}~\bibnamefont
  {Gover}}, \bibinfo {author} {\bibfnamefont {H.}~\bibnamefont {Zhang}},
  \bibinfo {author} {\bibfnamefont {T.}~\bibnamefont {Li}}, \bibinfo {author}
  {\bibfnamefont {L.}~\bibnamefont {Zhou}}, \ and\ \bibinfo {author}
  {\bibfnamefont {S.}~\bibnamefont {Zhu}},\ }\href {\doibase
  10.1103/PhysRevLett.122.173901} {\bibfield  {journal} {\bibinfo  {journal}
  {Phys. Rev. Lett.}\ }\textbf {\bibinfo {volume} {122}},\ \bibinfo {pages}
  {173901} (\bibinfo {year} {2019})}\BibitemShut {NoStop}%
\bibitem [{\citenamefont {Roushan}\ \emph {et~al.}(2017)\citenamefont
  {Roushan}, \citenamefont {Neill}, \citenamefont {Megrant}, \citenamefont
  {Chen}, \citenamefont {Babbush}, \citenamefont {Barends}, \citenamefont
  {Campbell}, \citenamefont {Chen}, \citenamefont {Chiaro}, \citenamefont
  {Dunsworth}, \citenamefont {Fowler}, \citenamefont {Jeffrey}, \citenamefont
  {Kelly}, \citenamefont {Lucero}, \citenamefont {Mutus}, \citenamefont
  {O'Malley}, \citenamefont {Neeley}, \citenamefont {Quintana}, \citenamefont
  {Sank}, \citenamefont {Vainsencher}, \citenamefont {Wenner}, \citenamefont
  {White}, \citenamefont {Kapit}, \citenamefont {Neven},\ and\ \citenamefont
  {Martinis}}]{58}%
  \BibitemOpen
  \bibfield  {author} {\bibinfo {author} {\bibfnamefont {P.}~\bibnamefont
  {Roushan}}, \bibinfo {author} {\bibfnamefont {C.}~\bibnamefont {Neill}},
  \bibinfo {author} {\bibfnamefont {A.}~\bibnamefont {Megrant}}, \bibinfo
  {author} {\bibfnamefont {Y.}~\bibnamefont {Chen}}, \bibinfo {author}
  {\bibfnamefont {R.}~\bibnamefont {Babbush}}, \bibinfo {author} {\bibfnamefont
  {R.}~\bibnamefont {Barends}}, \bibinfo {author} {\bibfnamefont
  {B.}~\bibnamefont {Campbell}}, \bibinfo {author} {\bibfnamefont
  {Z.}~\bibnamefont {Chen}}, \bibinfo {author} {\bibfnamefont {B.}~\bibnamefont
  {Chiaro}}, \bibinfo {author} {\bibfnamefont {A.}~\bibnamefont {Dunsworth}},
  \bibinfo {author} {\bibfnamefont {A.}~\bibnamefont {Fowler}}, \bibinfo
  {author} {\bibfnamefont {E.}~\bibnamefont {Jeffrey}}, \bibinfo {author}
  {\bibfnamefont {J.}~\bibnamefont {Kelly}}, \bibinfo {author} {\bibfnamefont
  {E.}~\bibnamefont {Lucero}}, \bibinfo {author} {\bibfnamefont
  {J.}~\bibnamefont {Mutus}}, \bibinfo {author} {\bibfnamefont
  {P.}~\bibnamefont {O'Malley}}, \bibinfo {author} {\bibfnamefont
  {M.}~\bibnamefont {Neeley}}, \bibinfo {author} {\bibfnamefont
  {C.}~\bibnamefont {Quintana}}, \bibinfo {author} {\bibfnamefont
  {D.}~\bibnamefont {Sank}}, \bibinfo {author} {\bibfnamefont {A.}~\bibnamefont
  {Vainsencher}}, \bibinfo {author} {\bibfnamefont {J.}~\bibnamefont {Wenner}},
  \bibinfo {author} {\bibfnamefont {T.}~\bibnamefont {White}}, \bibinfo
  {author} {\bibfnamefont {E.}~\bibnamefont {Kapit}}, \bibinfo {author}
  {\bibfnamefont {H.}~\bibnamefont {Neven}}, \ and\ \bibinfo {author}
  {\bibfnamefont {J.}~\bibnamefont {Martinis}},\ }\href {\doibase
  10.1038/NPHYS3930} {\bibfield  {journal} {\bibinfo  {journal} {NATURE
  PHYSICS}\ }\textbf {\bibinfo {volume} {13}},\ \bibinfo {pages} {146}
  (\bibinfo {year} {2017})}\BibitemShut {NoStop}%
\bibitem [{\citenamefont {Bomantara}\ and\ \citenamefont {Gong}(2020)}]{qubit}%
  \BibitemOpen
  \bibfield  {author} {\bibinfo {author} {\bibfnamefont {R.~W.}\ \bibnamefont
  {Bomantara}}\ and\ \bibinfo {author} {\bibfnamefont {J.}~\bibnamefont
  {Gong}},\ }\href {\doibase 10.1103/PhysRevB.101.085401} {\bibfield  {journal}
  {\bibinfo  {journal} {Phys. Rev. B}\ }\textbf {\bibinfo {volume} {101}},\
  \bibinfo {pages} {085401} (\bibinfo {year} {2020})}\BibitemShut {NoStop}%
\bibitem [{\citenamefont {McIver}\ \emph {et~al.}(2020)\citenamefont {McIver},
  \citenamefont {Schulte}, \citenamefont {Stein}, \citenamefont {Matsuyama},
  \citenamefont {Jotzu}, \citenamefont {Meier},\ and\ \citenamefont
  {Cavalleri}}]{59}%
  \BibitemOpen
  \bibfield  {author} {\bibinfo {author} {\bibfnamefont {J.}~\bibnamefont
  {McIver}}, \bibinfo {author} {\bibfnamefont {B.}~\bibnamefont {Schulte}},
  \bibinfo {author} {\bibfnamefont {F.}~\bibnamefont {Stein}}, \bibinfo
  {author} {\bibfnamefont {T.}~\bibnamefont {Matsuyama}}, \bibinfo {author}
  {\bibfnamefont {G.}~\bibnamefont {Jotzu}}, \bibinfo {author} {\bibfnamefont
  {G.}~\bibnamefont {Meier}}, \ and\ \bibinfo {author} {\bibfnamefont
  {A.}~\bibnamefont {Cavalleri}},\ }\href {\doibase 10.1038/s41567-019-0698-y}
  {\bibfield  {journal} {\bibinfo  {journal} {NATURE PHYSICS}\ }\textbf
  {\bibinfo {volume} {16}},\ \bibinfo {pages} {38} (\bibinfo {year}
  {2020})}\BibitemShut {NoStop}%
\bibitem [{\citenamefont {Cupo}\ \emph {et~al.}(2021)\citenamefont {Cupo},
  \citenamefont {Cobanera}, \citenamefont {Whitfield}, \citenamefont
  {Ramanathan},\ and\ \citenamefont {Viola}}]{gra1}%
  \BibitemOpen
  \bibfield  {author} {\bibinfo {author} {\bibfnamefont {A.}~\bibnamefont
  {Cupo}}, \bibinfo {author} {\bibfnamefont {E.}~\bibnamefont {Cobanera}},
  \bibinfo {author} {\bibfnamefont {J.~D.}\ \bibnamefont {Whitfield}}, \bibinfo
  {author} {\bibfnamefont {C.}~\bibnamefont {Ramanathan}}, \ and\ \bibinfo
  {author} {\bibfnamefont {L.}~\bibnamefont {Viola}},\ }\href {\doibase
  10.1103/PhysRevB.104.174304} {\bibfield  {journal} {\bibinfo  {journal}
  {Phys. Rev. B}\ }\textbf {\bibinfo {volume} {104}},\ \bibinfo {pages}
  {174304} (\bibinfo {year} {2021})}\BibitemShut {NoStop}%
\bibitem [{\citenamefont {Perez-Piskunow}\ \emph {et~al.}(2014)\citenamefont
  {Perez-Piskunow}, \citenamefont {Usaj}, \citenamefont {Balseiro},\ and\
  \citenamefont {Torres}}]{gra2}%
  \BibitemOpen
  \bibfield  {author} {\bibinfo {author} {\bibfnamefont {P.~M.}\ \bibnamefont
  {Perez-Piskunow}}, \bibinfo {author} {\bibfnamefont {G.}~\bibnamefont
  {Usaj}}, \bibinfo {author} {\bibfnamefont {C.~A.}\ \bibnamefont {Balseiro}},
  \ and\ \bibinfo {author} {\bibfnamefont {L.~E. F.~F.}\ \bibnamefont
  {Torres}},\ }\href {\doibase 10.1103/PhysRevB.89.121401} {\bibfield
  {journal} {\bibinfo  {journal} {Phys. Rev. B}\ }\textbf {\bibinfo {volume}
  {89}},\ \bibinfo {pages} {121401} (\bibinfo {year} {2014})}\BibitemShut
  {NoStop}%
\bibitem [{\citenamefont {Usaj}\ \emph {et~al.}(2014)\citenamefont {Usaj},
  \citenamefont {Perez-Piskunow}, \citenamefont {Foa~Torres},\ and\
  \citenamefont {Balseiro}}]{gra3}%
  \BibitemOpen
  \bibfield  {author} {\bibinfo {author} {\bibfnamefont {G.}~\bibnamefont
  {Usaj}}, \bibinfo {author} {\bibfnamefont {P.~M.}\ \bibnamefont
  {Perez-Piskunow}}, \bibinfo {author} {\bibfnamefont {L.~E.~F.}\ \bibnamefont
  {Foa~Torres}}, \ and\ \bibinfo {author} {\bibfnamefont {C.~A.}\ \bibnamefont
  {Balseiro}},\ }\href {\doibase 10.1103/PhysRevB.90.115423} {\bibfield
  {journal} {\bibinfo  {journal} {Phys. Rev. B}\ }\textbf {\bibinfo {volume}
  {90}},\ \bibinfo {pages} {115423} (\bibinfo {year} {2014})}\BibitemShut
  {NoStop}%
\bibitem [{\citenamefont {Wu}\ and\ \citenamefont {An}(2020)}]{an1}%
  \BibitemOpen
  \bibfield  {author} {\bibinfo {author} {\bibfnamefont {H.}~\bibnamefont
  {Wu}}\ and\ \bibinfo {author} {\bibfnamefont {J.-H.}\ \bibnamefont {An}},\
  }\href {\doibase 10.1103/PhysRevB.102.041119} {\bibfield  {journal} {\bibinfo
   {journal} {Phys. Rev. B}\ }\textbf {\bibinfo {volume} {102}},\ \bibinfo
  {pages} {041119} (\bibinfo {year} {2020})}\BibitemShut {NoStop}%
\bibitem [{\citenamefont {Wu}\ \emph {et~al.}(2021)\citenamefont {Wu},
  \citenamefont {Wang},\ and\ \citenamefont {An}}]{an2}%
  \BibitemOpen
  \bibfield  {author} {\bibinfo {author} {\bibfnamefont {H.}~\bibnamefont
  {Wu}}, \bibinfo {author} {\bibfnamefont {B.-Q.}\ \bibnamefont {Wang}}, \ and\
  \bibinfo {author} {\bibfnamefont {J.-H.}\ \bibnamefont {An}},\ }\href
  {\doibase 10.1103/PhysRevB.103.L041115} {\bibfield  {journal} {\bibinfo
  {journal} {Phys. Rev. B}\ }\textbf {\bibinfo {volume} {103}},\ \bibinfo
  {pages} {L041115} (\bibinfo {year} {2021})}\BibitemShut {NoStop}%
\bibitem [{\citenamefont {Zhou}\ \emph {et~al.}(2018)\citenamefont {Zhou},
  \citenamefont {Wang}, \citenamefont {Wang},\ and\ \citenamefont
  {Gong}}]{zhou1}%
  \BibitemOpen
  \bibfield  {author} {\bibinfo {author} {\bibfnamefont {L.}~\bibnamefont
  {Zhou}}, \bibinfo {author} {\bibfnamefont {Q.-h.}\ \bibnamefont {Wang}},
  \bibinfo {author} {\bibfnamefont {H.}~\bibnamefont {Wang}}, \ and\ \bibinfo
  {author} {\bibfnamefont {J.}~\bibnamefont {Gong}},\ }\href {\doibase
  10.1103/PhysRevA.98.022129} {\bibfield  {journal} {\bibinfo  {journal} {Phys.
  Rev. A}\ }\textbf {\bibinfo {volume} {98}},\ \bibinfo {pages} {022129}
  (\bibinfo {year} {2018})}\BibitemShut {NoStop}%
\bibitem [{\citenamefont {Bomantara}\ \emph {et~al.}(2016)\citenamefont
  {Bomantara}, \citenamefont {Raghava}, \citenamefont {Zhou},\ and\
  \citenamefont {Gong}}]{zhou2}%
  \BibitemOpen
  \bibfield  {author} {\bibinfo {author} {\bibfnamefont {R.~W.}\ \bibnamefont
  {Bomantara}}, \bibinfo {author} {\bibfnamefont {G.~N.}\ \bibnamefont
  {Raghava}}, \bibinfo {author} {\bibfnamefont {L.}~\bibnamefont {Zhou}}, \
  and\ \bibinfo {author} {\bibfnamefont {J.}~\bibnamefont {Gong}},\ }\href
  {\doibase 10.1103/PhysRevE.93.022209} {\bibfield  {journal} {\bibinfo
  {journal} {Phys. Rev. E}\ }\textbf {\bibinfo {volume} {93}},\ \bibinfo
  {pages} {022209} (\bibinfo {year} {2016})}\BibitemShut {NoStop}%
\bibitem [{\citenamefont {Rudner}\ \emph {et~al.}(2013)\citenamefont {Rudner},
  \citenamefont {Lindner}, \citenamefont {Berg},\ and\ \citenamefont
  {Levin}}]{60}%
  \BibitemOpen
  \bibfield  {author} {\bibinfo {author} {\bibfnamefont {M.~S.}\ \bibnamefont
  {Rudner}}, \bibinfo {author} {\bibfnamefont {N.~H.}\ \bibnamefont {Lindner}},
  \bibinfo {author} {\bibfnamefont {E.}~\bibnamefont {Berg}}, \ and\ \bibinfo
  {author} {\bibfnamefont {M.}~\bibnamefont {Levin}},\ }\href {\doibase
  10.1103/PhysRevX.3.031005} {\bibfield  {journal} {\bibinfo  {journal} {Phys.
  Rev. X}\ }\textbf {\bibinfo {volume} {3}},\ \bibinfo {pages} {031005}
  (\bibinfo {year} {2013})}\BibitemShut {NoStop}%
\bibitem [{\citenamefont {Cayssol}\ \emph {et~al.}(2013)\citenamefont
  {Cayssol}, \citenamefont {Dora}, \citenamefont {Simon},\ and\ \citenamefont
  {Moessner}}]{61}%
  \BibitemOpen
  \bibfield  {author} {\bibinfo {author} {\bibfnamefont {J.}~\bibnamefont
  {Cayssol}}, \bibinfo {author} {\bibfnamefont {B.}~\bibnamefont {Dora}},
  \bibinfo {author} {\bibfnamefont {F.}~\bibnamefont {Simon}}, \ and\ \bibinfo
  {author} {\bibfnamefont {R.}~\bibnamefont {Moessner}},\ }\href {\doibase
  10.1002/pssr.201206451} {\bibfield  {journal} {\bibinfo  {journal} {PHYSICA
  STATUS SOLIDI-RAPID RESEARCH LETTERS}\ }\textbf {\bibinfo {volume} {7}},\
  \bibinfo {pages} {101} (\bibinfo {year} {2013})}\BibitemShut {NoStop}%
\bibitem [{\citenamefont {Mukherjee}\ \emph {et~al.}(2017)\citenamefont
  {Mukherjee}, \citenamefont {Spracklen}, \citenamefont {Valiente},
  \citenamefont {Andersson}, \citenamefont {{\"O}hberg}, \citenamefont
  {Goldman},\ and\ \citenamefont {Thomson}}]{62}%
  \BibitemOpen
  \bibfield  {author} {\bibinfo {author} {\bibfnamefont {S.}~\bibnamefont
  {Mukherjee}}, \bibinfo {author} {\bibfnamefont {A.}~\bibnamefont
  {Spracklen}}, \bibinfo {author} {\bibfnamefont {M.}~\bibnamefont {Valiente}},
  \bibinfo {author} {\bibfnamefont {E.}~\bibnamefont {Andersson}}, \bibinfo
  {author} {\bibfnamefont {P.}~\bibnamefont {{\"O}hberg}}, \bibinfo {author}
  {\bibfnamefont {N.}~\bibnamefont {Goldman}}, \ and\ \bibinfo {author}
  {\bibfnamefont {R.~R.}\ \bibnamefont {Thomson}},\ }\href {\doibase
  10.1038/ncomms13918} {\bibfield  {journal} {\bibinfo  {journal} {Nature
  Communications}\ }\textbf {\bibinfo {volume} {8}},\ \bibinfo {pages} {13918}
  (\bibinfo {year} {2017})}\BibitemShut {NoStop}%
\bibitem [{\citenamefont {Zhou}\ and\ \citenamefont {Gong}(2018)}]{63}%
  \BibitemOpen
  \bibfield  {author} {\bibinfo {author} {\bibfnamefont {L.}~\bibnamefont
  {Zhou}}\ and\ \bibinfo {author} {\bibfnamefont {J.}~\bibnamefont {Gong}},\
  }\href {\doibase 10.1103/PhysRevB.98.205417} {\bibfield  {journal} {\bibinfo
  {journal} {Phys. Rev. B}\ }\textbf {\bibinfo {volume} {98}},\ \bibinfo
  {pages} {205417} (\bibinfo {year} {2018})}\BibitemShut {NoStop}%
\bibitem [{\citenamefont {Asb\'oth}(2012)}]{71}%
  \BibitemOpen
  \bibfield  {author} {\bibinfo {author} {\bibfnamefont {J.~K.}\ \bibnamefont
  {Asb\'oth}},\ }\href {\doibase 10.1103/PhysRevB.86.195414} {\bibfield
  {journal} {\bibinfo  {journal} {Phys. Rev. B}\ }\textbf {\bibinfo {volume}
  {86}},\ \bibinfo {pages} {195414} (\bibinfo {year} {2012})}\BibitemShut
  {NoStop}%
\bibitem [{\citenamefont {Asb\'oth}\ and\ \citenamefont {Obuse}(2013)}]{72}%
  \BibitemOpen
  \bibfield  {author} {\bibinfo {author} {\bibfnamefont {J.~K.}\ \bibnamefont
  {Asb\'oth}}\ and\ \bibinfo {author} {\bibfnamefont {H.}~\bibnamefont
  {Obuse}},\ }\href {\doibase 10.1103/PhysRevB.88.121406} {\bibfield  {journal}
  {\bibinfo  {journal} {Phys. Rev. B}\ }\textbf {\bibinfo {volume} {88}},\
  \bibinfo {pages} {121406} (\bibinfo {year} {2013})}\BibitemShut {NoStop}%
\bibitem [{\citenamefont {Yin}\ \emph {et~al.}(2018)\citenamefont {Yin},
  \citenamefont {Jiang}, \citenamefont {Li}, \citenamefont {L\"u},\ and\
  \citenamefont {Chen}}]{73}%
  \BibitemOpen
  \bibfield  {author} {\bibinfo {author} {\bibfnamefont {C.}~\bibnamefont
  {Yin}}, \bibinfo {author} {\bibfnamefont {H.}~\bibnamefont {Jiang}}, \bibinfo
  {author} {\bibfnamefont {L.}~\bibnamefont {Li}}, \bibinfo {author}
  {\bibfnamefont {R.}~\bibnamefont {L\"u}}, \ and\ \bibinfo {author}
  {\bibfnamefont {S.}~\bibnamefont {Chen}},\ }\href {\doibase
  10.1103/PhysRevA.97.052115} {\bibfield  {journal} {\bibinfo  {journal} {Phys.
  Rev. A}\ }\textbf {\bibinfo {volume} {97}},\ \bibinfo {pages} {052115}
  (\bibinfo {year} {2018})}\BibitemShut {NoStop}%
\bibitem [{\citenamefont {Groth}\ \emph {et~al.}(2014)\citenamefont {Groth},
  \citenamefont {Wimmer}, \citenamefont {Akhmerov},\ and\ \citenamefont
  {Waintal}}]{Gro74}%
  \BibitemOpen
  \bibfield  {author} {\bibinfo {author} {\bibfnamefont {C.~W.}\ \bibnamefont
  {Groth}}, \bibinfo {author} {\bibfnamefont {M.}~\bibnamefont {Wimmer}},
  \bibinfo {author} {\bibfnamefont {A.~R.}\ \bibnamefont {Akhmerov}}, \ and\
  \bibinfo {author} {\bibfnamefont {X.}~\bibnamefont {Waintal}},\ }\href
  {\doibase 10.1088/1367-2630/16/6/063065} {\bibfield  {journal} {\bibinfo
  {journal} {New Journal of Physics}\ }\textbf {\bibinfo {volume} {16}},\
  \bibinfo {pages} {063065} (\bibinfo {year} {2014})}\BibitemShut {NoStop}%
\end{thebibliography}%

\end{document}